\documentclass{article}
\usepackage[utf8]{inputenc}
\usepackage{graphicx}
\usepackage{amssymb}
\usepackage{lineno}
\usepackage{amsmath}
\usepackage{algorithmic}
\usepackage{algorithm}
\usepackage{multirow}
\usepackage{caption}
\usepackage{subcaption}

\usepackage{mathtools}

\usepackage{xcolor}

\newtheorem{definition}{Definition}[section]

\title{An Intelligent Edge-Centric Queries Allocation Scheme based on Ensemble Models}
\author{Kostas Kolomvatsos$^1$ \and Christos Anagnostopoulos$^2$}
\date{%
    $^1$Department of Informatics and Telecommunications, University of Thessaly, 35131 Lamia Greece\\%
    $^2$School of Computing Science, University of Glasgow, G12 8RZ Glasgow UK\\[2ex]%
}

\begin{document}

\maketitle

\begin{abstract}
The combination of Internet of Things (IoT) and 
Edge Computing (EC) can assist in the delivery of novel applications
that will facilitate end users activities.
Data collected by numerous devices present in the IoT 
infrastructure can be hosted into 
a set of EC nodes becoming the subject of processing tasks for the provision of
analytics.
Analytics are derived as the result of various queries defined 
by end users or applications.
Such queries can be executed in the available EC nodes
to limit the latency in the provision of responses.
In this paper, we propose a meta-ensemble learning scheme that supports the 
decision making for the allocation of queries to 
the appropriate EC nodes. 
Our learning model decides over queries' and nodes' characteristics.
We provide the description of a matching process between 
queries and nodes after concluding the contextual 
information for 
each envisioned characteristic adopted in our meta-ensemble scheme.
We rely on widely known ensemble models, combine them and offer an additional processing layer
to increase the performance.
The aim is to result a subset of EC nodes that will host
each incoming query.
Apart from the description of the proposed model,
we report on its evaluation
and the corresponding results. 
Through a large set of experiments and a numerical analysis, we aim at revealing the pros and 
cons of the proposed scheme.
\end{abstract}


\maketitle

\section{Introduction}
Currently, we are in the middle of a data management revolution brought 
by the numerous devices present in the \textit{Internet of Things} (IoT) infrastructure.
These devices are capable of interacting with their environment, 
perform some processing activities and report to the upper layers 
either the collected data or the produced knowledge.
On top of the discussed devices, 
the \textit{Edge Computing} (EC) structure,
the \textit{Fog layer} and, finally, the \textit{Cloud} 
provide multiple storage and processing locations where 
knowledge can be delivered
\cite{moysiadis}.
In this architecture, finding efficient techniques 
for data management becomes significant due to the 
high volumes of data produced by numerous devices
\cite{yongrui}.
Recent studies show that 
data offloading to the Cloud requires 100 to 200 ms of additional latency compared 
to EC technologies
\cite{chen}. 
Hence, to reduce the latency, we could store and process 
the collected data 
to EC nodes.
However, due to their limited computational capabilities
only 
a part of data can be stored and processed locally
while the 
remaining will be sent to Cloud.
Currently, local data processing
gains more attention as
any solution has to 
build over the limited delay in the provision of the requested analytics. 
Pushing as much computing workload 
as close to the edge as possible can 
bring serious benefits, particularly 
where communication costs are high 
or where instant action is needed \footnote{https://www.accenture.com/t20170628T011733Z\_\_w\_\_/us-en/\_acnmedia/PDF-54/Accenture-Insight-Mobility-Edge-Analytics.pdf}. 

\textit{Edge Nodes} (ENs)
can maintain a local dataset forming a network of distributed data repositories. 
Data may also be replicated for supporting 
fault tolerant applications.
Queries/tasks generated by end users or applications could request for analytics 
over the collected data.
As data are distributed among ENs, it is imperative to define a mechanism 
that allocates the incoming queries to the appropriate nodes. 
Such a decision should be made not only based on queries demand for data but also on various characteristics of the ENs themselves.

\textbf{\textit{Motivating Example}}. 
If a query asks for temperature recordings between 10 and 20 degrees in the Celsius scale, 
it is useless to allocate the query to a dataset where its statistics indicate 
temperature recordings above 20 degrees or below 10 degrees. 
Allocating the query to the specific dataset, we will waste resources and time for receiving an empty set.
In addition, when multiple datasets exhibit statistics 
`similar' to the incoming query, a load balancing aspect should be taken 
into consideration to avoid the creation of bottlenecks in the processing activities of each EN.

Traditionally, the demand for a low latency in the provision 
of responses is achieved by employing more computing resources and parallelizing the application logic over the datacenter infrastructure \cite{wen}.
In our case, the incoming queries can be distributed 
to the available ENs, however, 
such nodes are equipped 
with different resources and exhibit a 
different status.
It is worth noticing that a query/task could be allocated
in multiple ENs imposing new requirements
related to the aggregation of the 
final responses.
We should also keep in mind that 
the number of ENs is growing 
while the need for increased decentralization and autonomy 
is also imperative.
To reduce the latency in the provision of responses,
it is not enough to perform a parallel execution 
on top of the available ENs but also 
to allocate the incoming queries/tasks to 
appropriate ENs that will 
conclude the envisioned calculations in short time.
The turnaround time is affected by multiple parameters
like the load of ENs
(the higher the load is, the higher the response time becomes), the number and the statistics of data present in them
(the size of the datasets affects the response time)
and so on and so forth.
In this paper, we are motivated by  
the vast EC infrastructure where ENs exhibit a dynamic behaviour 
concluded by their interactions with IoT devices and the Cloud back end.
Any decision for offloading queries/tasks to the available ENs
should be mandated by their current status (i.e., their ability to start and conclude the execution in short time)
and the matching between the queries/tasks requirements 
(as dictated by the constraints imposed by end users of applications) and the data present in the EC infrastructure.
The baseline solution deals with a query/task  
allocated to all the available ENs which is governed by the 
following disadvantages:
(i) the requestor should wait to receive the aggregated results from all the available ENs. The aggregation process will start only when the
last EN reports its outcomes. Hence, the turnaround time is 
affected by the longest response time observed in
the infrastructure;
(ii) in the responses aggregation phase, (near) empty lists will be 
provided from ENs not having data related to queries/tasks constraints. Resources and time will be spent for processing such data that will, finally, will not add value 
to the overall processing while requiring time and resources to process them. 
We should not forget that ENs are usually nodes with limited 
computational and resources capabilities, thus, we should avoid unnecessary executions of queries/tasks.

In this paper, we propose an intelligent scheme that
takes into consideration the characteristics of both, 
queries and ENs to conclude the most efficient allocations.
Such characteristics are related to all aspects of queries and 
ENs and their matching, i.e., we rely on the `burden' that a  query adds to ENs, 
the data required by the query, the load and speed of ENs and so on and so forth.
We consider an entity, the \textit{Query Controller} (QC) 
that is responsible to receive queries (from this point forward, we refer
to queries when we want to depict the execution of a query/task) 
and proceed with their allocation. 
The QC can be present at the Cloud being directly connected with a set of 
\textit{Query Processors} (QPs) placed in every EN. 
QPs execute the incoming queries and return the final response to the interested QC.
Our scheme enhances the behaviour and the decision making capabilities of QCs.
We adopt the contextual information of queries and ENs characteristics 
at the time when the decision should be concluded.
We propose the use of multiple contextual vectors,
i.e., our basis for `matching' queries with the available nodes.
Our mechanism adopts a classification model over the 
the complexity class of any incoming query and
the estimation of the distance between the query constraints and the data present at every EN.
In the final decision making,
we use a meta-ensemble 
model that builds on top of multiple 
ensemble schemes.
This way, we build an `hierarchy' of classification models
adopted to deliver better results than individual classification modules.

The efficient allocation of queries to a number of nodes 
is also the subject of our previous efforts presented in \cite{kolomvatsos2}, \cite{kolomvatsos3} and \cite{kolomvatsos1}.
In \cite{kolomvatsos2}, every allocation is concluded
based on an optimal stopping scheme that examines all the available ENs 
concerning their ability to host and execute a query.
The model proposed in \cite{kolomvatsos1} aims at 
the adoption of a learning scheme that will, finally, decide
each allocation when a query arrives.
In \cite{kolomvatsos3}, we adopt reinforcement learning
and clustering to deliver the final allocation. 
Our motivation for extending our previous work is related to our desire to 
avoid the drawbacks of adopting a single solution.
For instance, when relying on a specific learning model,
we can meet the following disadvantages \cite{dietterich}:
(i) a single model can be viewed as searching a space for detecting the best possible hypothesis.
A statistical problem arises when the amount of training data is too small compared to the size of the hypotheses space;
(ii) many algorithms are trapped to local optima. Even if 
there are enough training data, it may be computational 
expensive to find the best hypothesis in the available space;
(iii) in most applications, the true function 
cannot be represented by any of the hypotheses in the search space.
We depart from our previous work and 
provide a scheme that avoids the drawbacks of individual models.
The contributions of this paper depict the differences from the previous efforts being 
described in the following list:
\begin{itemize}
	\item we propose a meta-ensemble learning scheme for queries allocation to a set of ENs. 
	The meta-ensemble learning scheme adopts multiple ensemble learning (sub-)models to provide a powerful and efficient allocation model.
	\item we propose a decision making mechanism that builds over the contextual data related to queries and ENs characteristics. Through the formulated contextual vectors, our decision making model is capable of `matching' the incoming queries with the appropriate ENs.
	\item we propose a model for delivering the complexity of any incoming query. The complexity class depicts 
	the `burden' that a query will add in ENs where it will be executed.
	\item we adopt a model for estimating the distance of a query with datasets present in ENs. 
	Such information is significant for the final allocation as we aim at allocating queries in datasets with which they exhibit the minimum distance. 
\end{itemize}

The paper is organized as follows.
Section \ref{section2} reports on the prior work in the field by presenting important activities related to our problem.
Section \ref{section3} presents the problem under consideration and some introductory information.
Section \ref{section4} discusses how to model the incoming queries and the envisioned ENs while Section \ref{section5} presents the proposed meta-ensemble learning scheme.
In Section \ref{section6}, we proceed with our experimental evaluation and in Section \ref{section7}, we conclude our paper by presenting our future research plans.

\section{Related Work}
\label{section2}
With the advent of IoT, numerous devices can interact with their environment and 
the network in order to collect and process data.
In many application domains, data play a central role
towards the generation and the provision of knowledge 
that will support the efficient decision making.
Example domains are financial services, life sciences, mobile services, etc.
Apart from the discussed devices, end users may also generate data  
like tweets, social networking interactions or photos
\cite{ref301}.
Through analytics, one can 
support the efficient decision making having a view on the
`meaning' of the collected data.
Analytics aim to discover patterns, especially in the case of 
unstructured data. 
Various tools for large scale data analytics have been proposed 
in the literature.
The majority of them concern 
batch oriented systems and they build over Hadoop
\cite{forrester1}.
A number of research efforts 
try to provide performance insights to 
the discussed framework \cite{ref303}, \cite{ref304}, \cite{ref305}.
Researchers provide new functionalities   
to increase the performance of legacy systems.
For instance, the authors of \cite{ref302} propose
the Starfish, a self-tuning tool for
large scale data analytics. 

The aforementioned processing can be either centralized or distributed, 
i.e., the processing takes place where data are initially collected.
We can also identify the need for streams processing
to facilitate the online, (near) real time provision of responses 
in a set of analytics queries. 
Queries/tasks allocation and scheduling are important research 
subjects in multiple domains.
Both subjects have a significant impact in the 
IoT and EC.
IoT/EC nodes have 
limited computational capabilities while being restricted 
by various energy constraints.
It seems that streams processing 
is the appropriate methodology for delivering real time analytics.
ENs can apply a decision making mechanism to
process the incoming tasks/queries.
They should take into consideration tasks'/queries' specific characteristics 
in combination with their current internal status 
for any decision making. 
Additionally, nodes 
may train and update machine learning models locally to serve the incoming 
tasks/queries.
These local models can be  
aggregated in an upper layer \cite{Daga}.
This approach is appropriate to detect local data shifts
and create collaborative learning and model-sharing environments in which 
local models can quickly adapt to any changes in the collected data. 

A widely studied research subject is task scheduling 
in \textit{Wireless Sensor Networks} (WSNs). 
Mapping and scheduling should 
take into consideration energy constraints to secure an efficient 
execution \cite{ref3}, \cite{ref16}.
A pre-processor and a scheduler can be responsible for the final allocation. 
The pre-processor tries to identify the energy requirements of the incoming tasks/queries and, 
based on energy monitoring activities, decides on the final scheduling. 
In any case, taking into consideration only a single parameter (e.g., remaining energy) in decision making can limit the reasoning capabilities.
Another approach is to study a fair energy balance among sensors 
while minimizing the delay using a market-based architecture \cite{edalat}. 
Taking into consideration the defined constraints, nodes may cooperate 
to conclude the final allocations \cite{ref2}.
Example algorithms involve tasks/queries clustering and node assignment mechanisms 
based on tasks/queries duplication and migration schemes. 
The aim is to minimize the execution time, thus, 
to deliver the final response with minimized latency. 
A model that could be adopted for such purposes is to cluster the 
network and build intra-cluster and inter-cluster scheduling relations. 
An \textit{Integer Linear Programming} (ILP) formulation and a 3-phase heuristic are also adopted to 
solve the allocation problem in 
\cite{yu}.
In \cite{yang}, the authors propose a modified
version of the binary \textit{Particle Swarm Optimization} (PSO).
The method adopts a different transfer function, a new position updating
process and mutation for the
task/query allocation problem.  
Another PSO-based solution is presented in \cite{razavinegad} which allocates tasks/queries 
into a number of robots trying to decrease the communication cost.
It is worth noticing that PSO-based models suffer from a low convergence rate while
they can be trapped to a local optimal especially in complex problems.
In \cite{hu}, the authors present a task/query allocation mechanism of a 
dynamic alliance based on a Genetic Algorithm to acquire the balance between energy consumption and accuracy. 
In \cite{coltin}, the authors discuss three algorithms to solve the discussed problem: 
a centralized, an auction-based, and a distributed algorithm. 
The distributed algorithm adopts a spanning tree over the static sensors to assign tasks/queries.

The continuous reporting 
of queries demanding for immediate responses is also 
the subject of various research efforts.
When large scale data applications involve 
continuous queries, for having
near real-time responses,
such applications, usually, involve a large number of 
data partitions.
Obtaining a response in near real-time
could be very difficult due to limitations defined 
by the amount of data and the underlying hardware performance.
Querying data samples and the 
provision of progressive analytics is
an efficient solution for the described problem \cite{ref201}.
Specific sampling techniques have been proposed 
\cite{ref310}, \cite{ref306}.
In progressive analytics, the
\textit{Approximate Query Processing} (AQP) technique can secure the accuracy in 
early results and
provide the corresponding confidence 
\cite{ref310}, \cite{ref314}, \cite{ref315}. 
Users defining queries are not involved in the process,
however, based on the retrieved confidence, they could rely on
an intelligent mechanism for 
handling early results.
When accuracy is at acceptable levels (according the specific
application domain), the process could be stopped. 

An analytics provisioning system is presented in 
\cite{refm1}. The system is based on the Prism framework 
and allows users to communicate 
samples to the system.
Queries are processed over the defined samples.
The authors propose Now!, a progressive
data-parallel computation framework for 
Windows Azure.
Now! mainly works with streaming engines to 
support progressive SQL over large scale
data. 
It is worth noticing that the selection of samples affects the 
statistical error of the final (sub-)dataset, thus, it may 
have negative consequences in the final decision making.
In \cite{ref308}, the authors present an online MapReduce 
scheme that supports online aggregation and continuous queries. 
For decreasing the latency of the system, the authors
propose to have the Map task sending early results
to the Reduce tasks. 
This mechanism enables the generation of approximate results, 
which is
particularly useful for interactive analytics.
In \cite{ref309}, the authors present a continuous
MapReduce model. The execution of Map and Reduce functions
is coordinated by a data stream processing platform. 
Latency is improved through a model where mappers are 
continually fed by data and the 
retrieved results are transferred to reducers. 
CONTROL \cite{ref306} is an AQP system intended to support
progressive analytics. 
Users have the opportunity to refine answers 
and have online control
of processing.  
DBO \cite{ref307} is another AQP system capable of calculating 
the exact answer to queries
over a large relational database. 
DBO can have an insight on the final response
together with specific bounds for the accuracy of early 
results.
As more information is processed, the DBO can provide provide more accurate results. 
Users can stop the process at any time, if the accuracy level is 
sufficient.

\section{Rationale \& Preliminaries}
\label{section3}
In this section, we provide a description of the interacting entities
towards the efficient allocation of the incoming queries
and present the problem under consideration.
We also setup the basis for 
solving the problem and apply the envisioned 
meta-ensemble classification scheme.

\subsection{Edge Network Architecture}
\begin{figure}[h]
\centering
\includegraphics[width=0.50\textwidth]{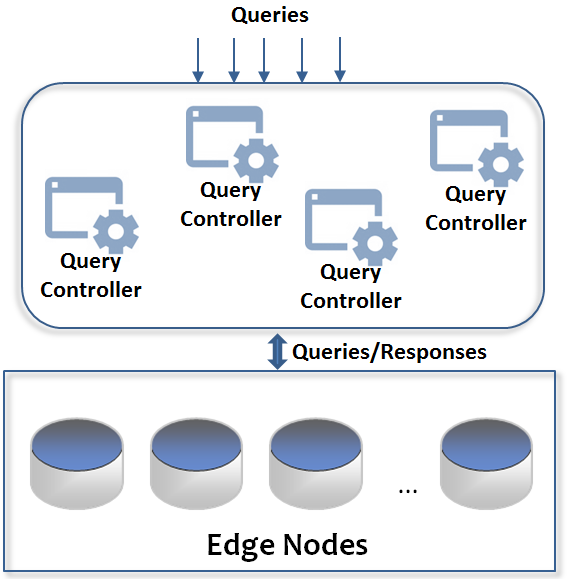}
\caption{The connection of query controllers and edge nodes.}
\label{fig1}
\end{figure}

Consider the setting presented in Figure \ref{fig1} where
$N$ ENs collect contextual data and locally 
process them in light of producing knowledge; 
each EN is indexed in the set $\mathcal{N} = \left\lbrace n_{1}, n_{2}, \ldots, n_{N} \right\rbrace$.
Contextual data are either collected by various sensors/end devices (e.g., users' smartphones) or are generated through local processing. 

\begin{definition}
Contextual data are depicted by the information that provides context (a value for a specific attribute) 
to an event, entity or a processing activity. 
\end{definition}

Every EN has specific characteristics related to its computational 
power and a limited storage capacity. 
Hence, only a subset of contextual data could be stored locally. 
The remaining data are sent to the back end infrastructure present at Cloud.
In this context, by supporting local data processing at the EN, it facilitates the minimization of latency in the provision of responses.

Local data processing involves statistical reasoning, inferential analytics, and real-time data management, e.g., estimation of the top-$k$ lists over the incoming data streams \cite{kolomvatsostopk}).
A dataset $DS_{i}$ is available over which the local data processing takes place. 
This dataset is continuously updated as fresh data arrive through streams.
The $i$th EN stores multivariate vectors in $DS_{i}$,
i.e., 
$\mathbf{x} = [x_{1}, x_{2}, \ldots, x_{L}]^{\top} \in \mathbb{R}^{L}$ ($L$ is the number of
dimensions - contextual attributes). 
As multiple devices may report vectorial data to multiple ENs, 
replicates among $DS_{i}$ and $DS_{j}$, with $i \neq j$ may be present. 
The management of potential replicas is beyond the scope of this paper. 
The whole data at the network edge form the set $\mathcal{DS} = \{DS_{1}, \ldots, DS_{N}\}$. 
In each EN, we introduce a local QP responsible to (i) receive a stream of 
analytics queries (e.g., estimating the regression plane among contextual variables within a time frame) 
from end users (e.g., applications, data analysts); (ii) execute them over $DS_{i}$ and 
(iii) send the results back to the requestor. 

\begin{definition}
An analytics query is a request for information 
responded by meaningful patterns found in the available dataset while being extracted based on a scientific process.
\end{definition}

We associate the $i$th QP with the $i$th EN, thus, 
we have $N$ ENs/QPs in the set $\mathcal{QP} = \left\lbrace qp_{1}, qp_{2}, \ldots, qp_{N} \right\rbrace$. 
We adopt a queue ion every QP which can handle a maximum number of queries. 
Without loss of generality, we consider that queues adopted in QPs have the same length.
Each QP has specific characteristics encoded in the set 
$\mathcal{C}_{i} = \left\lbrace c_{i,1}, c_{i,2}, \ldots, c_{i,m} \right\rbrace$. 
For instance, $\mathcal{C}_{i} = \left\lbrace l, s \right\rbrace$ with $l$ representing the current load 
and $s$ depicting the speed of the corresponding QP. 

\begin{definition}
The load is the quantity of analytics queries that can be carried at one time by an EN.
\end{definition}

\begin{definition}
The processing speed of an EN is its rate of execution of analytics queries.
\end{definition}

$l$ can be easily estimated through the current number of queries waiting in the corresponding queue, 
while $s$ indicates the throughput of a QP, i.e., the number of queries responded in a given time unit.

In the upper layer, shown in Figure \ref{fig1} (i.e., Fog/Cloud), there is 
a federation of QCs. 
QCs play the role of endpoints for end users or applications 
and achieve efficient provision of responses to incoming queries. 
QCs have direct access to the ENs, thus, to their corresponding QPs and they interact with them 
to get responses for the incoming queries. 
QCs, after receiving a query, should be able to find the most appropriate subset of QPs for 
allocating the query. 
Afterwards, QCs should aggregate the partial results 
to derive the final aggregated response, which will be delivered to end-users/applications. 
With the term \textit{appropriate}, we 
refer to the subset of those QPs that, at the specific time the query is issued, 
they exhibit characteristics that will facilitate its 
\textit{efficient execution} and return the result in the minimum turn around time.
The efficient execution is realized through the realization of various parameters that should 
be optimal for any allocation.
For instance, the response time should be limited, the 
outcomes should match against the query constraints, the response time
should fulfil the pre-defined time constraints (if set) and so on and so forth.
It should be noted that QCs, through their continuous interaction with the ENs and their QPs, 
can maintain historical performance data as well as the statistics of data present in each EN. 
Based on this context, QCs obtain a holistic overview on the current status of QPs and the data that 
they are stored locally in ENs. 

The incoming queries are represented via a stream: 
$Q = \left\lbrace q_{1}, q_{2}, \ldots q_{j} \right\rbrace$.
At time instance $t \in \mathbb{T} = \{1, 2, \ldots\}$ a query $q_{t}$ arrives to a QC. 
$q_{t}$ belongs to a specific \textit{query class}, 
exhibiting specific characteristics, i.e., 
$C^{q} = \left\lbrace c_{1}^{q}, c_{2}^{q}, \ldots, c_{|C^{q}|}^{q} \right\rbrace$.
Suppose that all queries have the same number and types of characteristics. 
For instance, $C^{q} = \left\lbrace p, a \right\rbrace$ where $p$ stands for the computational complexity and 
$a$ depicts a deadline for delivering the final result.

\begin{definition}
The computational complexity of an analytics query represents the 
amount of the required resources to execute it.
\end{definition}

If we focus on a database management system, we can identify the following query classes:
(i) \textit{Selection Queries}: The main representative of such queries is the SELECT command. This type of queries aim to provide data (tuples) that fulfil specific conditions; 
(ii) \textit{Modification Queries}: Such queries are adopted to perform changes/modifications in the underlying data, e.g., UPDATE. 
In most of the cases, these queries are demanding and require 
significant computational time and resources;
(iii) \textit{Aggregation Queries}: Usually, they are executed over other queries and apply algebraic operators, e.g.,  AVG, SUM, MIN, MAX, etc.
According to \cite{Hameurlain}, the generic characteristics of queries are: (i) the \textit{type} of the query, e.g., repetitive, ad-hoc; (ii) the query \textit{shape}; and (iii) the \textit{size} of the query, e.g., simple, medium, complex. 
Based on the characteristics of each query class, 
specific execution plans could be defined in the form of processing trees \cite{Hameurlain}. 
In \cite{Vashistha}, one can find a study on the calculation 
of queries complexity, however, with the focus being in the underlying databases.
In the database community, the complexity of a query is, usually, measured 
through the resources required by the database server for executing it. 
The most significant parameters for doing that is to focus on the space and time required for executing the query during the optimization phase. 
The query optimizer compares different execution strategies and selects the one with the least expected cost (or the maximum expected reward).
In our context, we want to adopt additional parameters not being bounded 
to the internal processes of any data management system but being 
aligned with the needs of an EC setting.
We focus on the upper layer of the aforementioned architecture and try to 
limit the time required for initiating the execution of a query.
We extend the aforementioned characteristics' list and incorporate new, `high level', parameters that depict the complexity and the need for instant response. 

\subsection{Problem Definition}
For the description of our problem
let us focus on an individual QC attached to a stream of queries. 
Suppose that at time $t$, a query $q_{t}$ arrives 
at the QC demanding for immediate processing. The QC should be based on $q_{t}$'s and ENs 
characteristics to take the appropriate decision and find the best 
subset of nodes to engage for the execution of $q_{t}$.
Among others,
we are interested in data constraints as dictated by $q_{t}$.
For instance, 
in case of an SQL query, we focus on the WHERE clause. 
These constraints define the conditions that should be met when 
we retrieve data to construct the final response for $q_{t}$.
To easily depict the discussed constraints,
we represent $q_{t}$ as a $2L$-dimensional vector 
\begin{eqnarray}
\mathbf{w} = [ \left\lbrace min_{1}, max_{1} \right\rbrace, \ldots, \left\lbrace min_{L}, max_{L} \right\rbrace ]^{\top} \in \mathbb{R}^{2L}
\end{eqnarray}
such that $\left\lbrace min_{i}, max_{i} \right\rbrace$
are the minimum and the maximum values defined 
for the $i$-th dimension (attribute).
Constraints should be matched against data present in ENs 
where, potentially, $q_{t}$ is directed to be executed. 
After the reception of $q_{t}$, the QC creates $N$
\textit{context vectors} (one for each QP) 
referring 
to $q_{t}$'s characteristics and 
the current status of ENs/QPs.
Context vectors have the following form:
\begin{eqnarray}
\mathbf{v}_{i} = \left\langle o, a, r_{i}, l_{i}, s_{i}\right\rangle
\end{eqnarray}
In addition, 
$o$ is the $q_{t}$'s expected complexity (elaborated later), 
$a$ is the deadline set for $q_{t}$, 
$r_{i}$ is the \textit{information relevance} of $q_{t}$
with $DS_{i}$ (the dataset present at the $i$th EN) and 
$l_{i}$ and $s_{i}$
are the load and the speed of the $i$th QP, respectively.
The aforementioned vectors represent the 
minimum sufficient statistics for both an incoming query and each EN/QP based on which the 
QC should decide the final allocation.
It should be noted that context vectors are easily concluded in short time
either using: (i) pre-defined values (e.g., $a$ is set during the reception of 
$q_{t}$, $s{i}$ is defined for each EN beforehand - it depends on the 
internal characteristics of QPs); 
(ii) simple calculations (e.g., $l_{i}$ can be extracted by the number of queries waiting for processing in the corresponding queue); or 
(iii) our proposed models (e.g., for concluding $o$ and $r_{i}$).
Overall, context vectors are fed into our proposed ensemble 
scheme for predicting the appropriateness of each EN/QP to be the host of each query. 

\textbf{Problem:}
\textit{
Given an incoming query $q_{t}$ represented by $\mathbf{w}$ and the associated context vectors 
$\{\mathbf{v}\}_{i=1}^{N}$, predict the most appropriate subset of QPs that should be engaged for the execution of that query.
}

In the remainder, we elaborate on the 
creation of the context vectors. 
It is worth mentioning that QCs can have:
(i) access to the statistical synopses (digests) of data present in ENs
(i.e., the QC exploits the minimum sufficient statistics of each dataset); (ii) access to statistical data related to the performance of ENs.
Based on the above, our mechanism is, both, \textit{performance-aware} 
and \textit{data-aware}.
We deal with the estimation of the ability of a QP to 
efficiently execute a query in the minimum time while 
delivering the outcomes that perfectly match to 
queries' constraints.

\section{Processors Characteristics and Queries Complexity}
\label{section4}

\subsection{Query Processor Speed \& Load}
As noted, ENs maintain a queue where the incoming queries wait for processing. 
The
size of the queue is adopted such that it can deliver up to $l_{i}$ 
queries per time unit, i.e., the percentage of the maximum load that can be afforded by the $i$th EN. 
Without loss of generality, we get $l_{i} \in [0,1]$
given that a maximum queue size is adopted for such purposes. 
When $l_{i} \to 1$, the EN experiences a high load. 
The load is also connected with the throughput and the queries 
reporting rate. 
Additionally, $s_{i}$ depicts the speed of processing. 
A resource demanding query, e.g., a join query, 
potentially requires more execution time and resources compared with 
a less resource demanding query like simple select commands. 
$s_{i}$ is, then, directly affecting the throughput, 
i.e., the number of queries executed 
in a time unit and assumes values in $[0,1]$. 
When $s_{i} \to 1$, the $i$th EN 
exhibits the best possible speed
approaching the maximum theoretical performance.
When $s_{i} \to 0$, the $i$th EN delays in the delivery responses to the the incoming queries.

\subsection{Query Complexity}
The classification of the computational complexity $o$ of $q_{t}$ and its effect on
$l$ and $s$ is the key element in our scheme. 
Various research efforts study the complexity of queries \cite{Artail}, \cite{Simon}, \cite{Vashistha}.
For realizing $o$, we rely on our previous effort 
presented in \cite{kolomvatsos4}.
We focus on a simple,
however, efficient mechanism that will deliver the final outcome
in the minimum time.
We consider that 
$|\Theta|$ complexity classes are available,
i.e., $\Theta = \left\lbrace \theta_{1}, \theta_{2}, \ldots, \theta_{|\Theta|} \right\rbrace$.
$\theta_{i}$ is aligned with the complexity performed by the operations of $q_{t}$ 
as required for producing the final result. 
We assign $q_{t}$ to a complexity class $\theta$ 
based on a typical classification task. 
The final complexity is defined based not only on quantitative characteristics, 
e.g., number of constraints and conditions, but also on qualitative characteristics, 
e.g., type of operations. 
For handling this complicated process, 
we introduce a fuzzy logic based approach and define a 
\textit{Fuzzy Classification Process} (FCP). 
The FCP evaluates the membership of $q_{t}$ 
in each of the pre-defined complexity classes.
Hence,  
we can obtain an estimate of the computational burden added to a candidate EN.
The FCP is executed over a set of historical executed queries 
along with their corresponding classes.
This means that QCs maintain the previous queries set 
realized by past interactions with ENs being dynamically updated 
as new requests for processing arrive.
The aforementioned set 
can be concluded by servers' logs reported after each processing
activity.
In any case, the creation of the historical set of queries and
the pairing process with the available complexity classes
is beyond of the scope of this work. 
We deliver 
a scheme that can be adopted in 
processing of 
real, unknown queries
providing the final result in limited time.
When focusing on a query decomposition process (it is the theoretical ground
coming from the database community)
to calculate the complexity, we need 
from 3.0 to 22.65 seconds depending on the query and the 
underlying DBMS \cite{lubke}.
This amount of time should be added to the time required for the
allocation of a query and the waiting time in the queue of the selected EN.
In our work, we specifically focus on the minimization of the allocation time
together with the reduction of the waiting time.  
We try to limit the time 
at every part of the envisioned processing, i.e., the 
initial allocation of $q_{t}$ combined with models
proposed by the database community aiming at query execution optimization. 

For evaluating $\theta$ for every $q_{t}$, 
we adopt 
widely known
similarity techniques instead of 
relying on a machine learning model
that requires a training phase. 
We try to minimize the 
complexity of our scheme,
proposing the use of a simple, fast, however, efficient 
model that delivers the final outcome in the minimum time.
Let the available dataset 
of training queries
be $DS_{Q}$ 
containing
tuples in the form
$\left\langle p_{k}, \theta_{k} \right\rangle$,
$k \in \left\lbrace 1, 2, \ldots, |DS_{Q}|\right\rbrace$.
$p_{k}$ represents $q_{t}$'s
statement along with its complexity class $\theta_{k} \in \Theta$. 
An example statement could be: 
$p_{k} = $ \texttt{SELECT NAME, PRICE FROM STOCKS WHERE PRICE < = 100 AND PRICE >= 10}. 
We, then, develop a function $f$ over $q_{t}$ and, 
based on $DS_{Q}$, deliver a vector encoding 
the similarity of $q_{t}$ with every complexity class in $\Theta$: 
\begin{eqnarray}
f(q; D_{Q}) \to \mathbf{q}^{s} \in [0,1]^{|\Theta|}
\end{eqnarray}
$\mathbf{q}^{s}$'s components
assume values in [0,1] 
demonstrating the degree of membership of 
$q_{t}$ to each complexity class, 
thus, forming the basis of our FCP. 
For instance, consider the vector $\mathbf{q}^{s} = [0.2, 0.8, 0.3]^{\top}$ given the complexity classes: $\Theta = \left\lbrace \theta_{1}=O(n \log n), \theta_{2} = O(n), \theta_{3} = O(n^{2}) \right\rbrace$. 
In this example, $\mathbf{q}^{s}$
shows that $q_{t}$ belongs with 20\% to 
$\theta_{1}$, 80\% to $\theta_{2}$ and 30\% to $\theta_{3}$. 

\begin{figure}[h]
\centering
\includegraphics[width=0.8\textwidth]{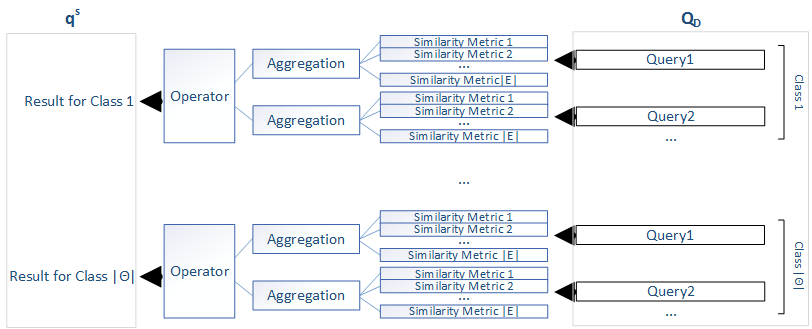}
\caption{The envisioned similarity aggregation process.}
\label{aggregation}
\end{figure}

For calculating 
every 
component of $\mathbf{q}^{s}$, 
we 
adopt a set of metrics
that deliver the similarity between vectors. 
The interested reader
can refer in \cite{kul} for more details. 
We propose an ensemble scheme for evaluating the 
similarity of $q_{t}$ with 
every tuple $\left\langle p_{k}, \theta_{k} \right\rangle \in DS_{Q}$.
Recall that tuples present in $DS_{Q}$ depict a set of training queries along 
with their complexity classes.
Our aim is to find the similarity of $q_{t}$ with every training query
merging the results in a subsequent step to deliver 
the final $q_{t}$'s
similarity with each of the pre-defined 
complexity classes.
The main theoretical foundation behind 
the adoption of an ensemble similarity scheme
is related to the special requirements of the problem and the need for avoiding 
any wrong decisions when relying on a single metric.
The ensemble scheme `combines' the `opinions'
of multiple metrics while the aggregated result is the one that will support the final outcome. 
Our ensemble scheme utilizes a set 
$E = \left\lbrace e_{1}, e_{2}, \ldots, e_{|E|} \right\rbrace$
of similarity metrics. 
$E$ can involve the Hamming distance \cite{pandit}, 
the Jaccard coefficient \cite{moulton}, 
and the Cosine similarity
\cite{pandit}. 
Any distance metric available in the
literature could be transformed to depict the
similarity between $q_{t}$ and the $p_{k}$ 
from the tuple $\left\langle p_{k}, \theta_{k} \right\rangle$. 
For instance, if $e_{d}$ is the Euclidean distance between $q_{t}$ and $p_{k}$, their similarity can
be defined as $\frac{1}{1+e_{d}}$.
The adopted similarity metrics are applied on each tuple `pre-classified' 
to $\theta_{k}$ aggregated to determine the 
$k$-th component $q^{s}_{k}$ 
of the complexity similarity vector.
Formally the `2D aggregation' (see Figure \ref{aggregation} - retrieved by \cite{kolomvatsos4}) is calculated as follows: 
$q^{s}_{k} = \Omega(\omega\left\lbrace e_{i}(q_{t},\left\langle s_{k}, \theta_{k} \right\rangle) \right\rbrace, \forall
 i$, $\forall \left\langle p_{k}, \theta_{k} \right\rangle$.
$\omega$ 
realizes the envisioned ensemble similarity scheme while the aggregation operator
$\Omega$ produces the $q^{s}_{k}$ 
over
multiple 
$\omega$ values.

For $\omega$, we consider that every 
individual
result (i.e., $e_{i}(q_{t},\left\langle p_{k}, \theta_{k} \right\rangle)$ represents
the membership of $q_{t}$ to a `virtual'
fuzzy set.
We have $|\mathcal{E}|$ membership degrees combined
to get the final similarity for each 
individual
tuple.
For instance, if we get $e_{1}=0.2$, $e_{2}=0.5$ and 
$e_{3}=0.3$, $q_{t}$ 
`belongs' to the $e_{1}$ fuzzy set by 0.2,
to the $e_{2}$ by 0.5 and 
to the $e_{3}$ by 0.3.
$\omega$ is a \textit{fuzzy aggregation operator}; 
an $|\mathcal{E}|$-place function $\omega:[0,1]^{|\mathcal{E}|} \to [0,1]$) that takes 
into consideration the membership to every fuzzy set and returns 
the final value.
Aggregation operators
are well studied in various research efforts \cite{farahbod}, \cite{hossain}.
Through a high number of experiments \cite{farahbod}, \cite{hossain}, 
the Einstein product, the algebraic product, 
the Hamacher product \cite{hossain} and the Schweizer-Sklar metric \cite{farahbod} are 
identified to have the best performance. 
In the proposed model, we adopt the Hamacher product as it gives us more opportunities
to `tune' the result through the parameter $\alpha \geq 0$.
The final $\omega$ value is defined as:
\[
\omega = \frac{\dot{e}\cdot \ddot{e}}{a + (1 - a)(\dot{e} + \ddot{e} - \dot{e}\cdot \ddot{e})}
\]
where $\dot{e}$ and $\ddot{e}$ are two similarity values.
As 
similarity metrics may `disagree', we propose the use of 
the top-$n$ similarity values based on their
significance level.
The \textit{Significance Level} (SL) depicts if a 
similarity
value is `representative' 
for many other outcomes.
We borrow the idea from the 
density based clustering \cite{han} where
centroids of the detected clusters are points
that `attract' many other data features around them in close distance.
We propose the use of the radius $\gamma$ and calculate the SL
for each similarity result
as follows:
$SL_{e_{i}} = \frac{1}{1+e^{-\left( \delta_{1} |d(e_{i}, e_{k})\leq \gamma| - \delta_{2} \right)}}, \forall i$,
where $\delta_{1}$ and $\delta_{2}$ are parameters adopted to smooth the sigmoid function.
With the sigmoid function, we eliminate the SL of 
similarity
values with 
a low number of `neighbors' in the radius $\gamma$.
The final results are sorted in a descending order of the SL and the top-$n$
of them are processed with the Hamacher product to deliver the final 
$\omega$.

The $\Omega$ operator builds over 
$\omega$ values produced for each tuple in $Q_{D}$
classified in $\theta_{k}$.
Let $\omega_{1}, \omega_{2}, \ldots, \omega_{m}$ are those
values.
For their aggregation, we rely
on a Quasi-Arithmetic mean, 
i.e.,
$q^{s}_{k} = \left[ \frac{1}{m} \sum_{i=1}^{m} \omega_{i}^{\alpha} \right]^{\frac{1}{\alpha}}$
where $\alpha$ is a parameter that `tunes' the function. 
When $\alpha = 1$, the function is the arithmetic mean,
when $\alpha = 2$, it is the quadratic mean and so on
and so forth.
After calculating the final values for each $\theta_{k}$
(i.e., realized by $\Omega_{k}$), we get $\mathbf{q^{s}} = \left\langle \Omega_{1}, \Omega_{2}, \ldots, \Omega_{|\Theta|} \right\rangle$.

\subsection{Distance Between Queries and Datasets}
We propose a distance model aiming at concluding 
$r_{i}$ adopted in the context vector 
$\mathbf{v}$.
Recall that 
$r_{i}$ is adopted to depict the similarity between 
$\mathbf{w}$ and data present 
in ENs.
We consider that the dimensions of the collected/stored vectors 
are not correlated.
At pre-defined intervals, ENs send to 
QCs the
statistics of local
data 
expressed by
two vectors,
i.e.,
the vector of means $\mathbf{\mu} = \left\langle \mu_{1}, \mu_{2}, \ldots,
\mu_{L} \right\rangle$ and the 
vector containing the standard deviation for 
each dimension, i.e.,
$\mathbf{\sigma} = \left\langle \sigma_{1}, \sigma_{2}, \ldots,
\sigma_{L} \right\rangle$.
Again, we have to 
recall that $\mathbf{w}$ represents the intervals
depicted by the constraints of $q_{t}$.
Our intention is to find the overlapping
between 
two sets of intervals, i.e.,
intervals defined by $\mathbf{w}$ and intervals
represented by data statistics 
(the combination of 
$\mathbf{\mu}$ and $\mathbf{\sigma}$).
Actually, $\mathbf{\mu}$ and $\mathbf{\sigma}$ can define the 
confidence interval 
for each dataset as follows.
We have 
$\mathbf{\mu} \pm z \cdot \frac{\mathbf{\sigma}}{|DS_{i}|}$
for all the adopted dimensions with $|DS_{i}|$ depicting the cardinality
of the corresponding dataset. 
$z$ represents the 
z-value retrieved by the standard normal (Z-) distribution 
for our desired confidence level.
The area between $-z$ and $+z$ is, approximately, the confidence percentage. 
For instance, for $z = 1.28$, the area between $-1.28$ and $+1.28$ is, approximately, 0.80. 

Based on the above
analysis, we 
proceed to 
calculations related to
the similarity between 
$\mathbf{w}$ and $N$ vectors,
i.e., 
\[
\mathbf{f}_{i} = \left\langle \left\lbrace \mu_{i1} - z \cdot \sigma_{i1}, \mu_{i1} + z \cdot \sigma_{i1}\right\rbrace, \ldots, 
 \left\lbrace \mu_{iL} - z \cdot \sigma_{iL}, \mu_{iL} + z \cdot \sigma_{iL}\right\rbrace \right\rangle, \forall i
\]
\footnote{We consider that every $\mathbf{\sigma}_{i}$ vector depicts the result $z \cdot \frac{\mathbf{\sigma}}{|DS_{i}|}$}.
This way, we conclude $N$ similarity values, one for each dataset, thus,
we deliver $N$ context vectors $\mathbf{v}$.
For deriving the final $r_{i}$,
we have to find the final similarity between
$L$ intervals, 
i.e., $\mathbf{w}$ and $\mathbf{f}_{i}$.
Typical distance/similarity measures (e.g., the Euclidean distance)
cannot efficiently manage cases where $\mathbf{w}$
can be completely contained in 
$\mathbf{f}_{i}$.
We rely on 
the research performed in
\cite{habler},
where a study on calculating the distance 
over interval data 
is provided. 
Based on the discussed metrics, we propose
the use of the overlapping metric $\psi_{k}$
to finally deliver
$r_{i}$ as follows:
$r_{i}\left(\mathbf{w}, \mathbf{f}_{i} \right) = h(\psi_{ik}), \forall
k \in \left\lbrace 1,2, \ldots, L\right\rbrace$.
In this equation, we propose the use of the aggregation function $h$
responsible to aggregate $L$
distance results.
Our calculations are exposed by the following equation
(applied for each dimension of the adopted dataset):
$\psi_{ik} = 1 - \frac{\lVert \mathbf{w} \cap \mathbf{f}_{i} \rVert}{\min \left\lbrace \lVert min_{k}, \max_{k} \rVert, \lVert \mu_{ik}-\sigma_{ik}, \mu_{ik}+\sigma_{ik} \rVert \right\rbrace}$ where
\begin{eqnarray}
\resizebox{1.0\hsize}{!}{

$w_{k} \cap f_{ik} = \left\{
	\begin{array}{ll}
		\left(\max\left\lbrace min_{k}, \mu_{ik}-\sigma_{ik} \right\rbrace, \min \left\lbrace \max_{k}, \mu_{ik}+\sigma_{ik} \right\rbrace \right)  & \textit{for }\max\left\lbrace \min_{k}, \mu_{ik}-\sigma_{ik} \right\rbrace < \min \left\lbrace \max_{k}, \mu_{ik}+\sigma_{ik} \right\rbrace \\
		0 & \mbox{Otherwise }
	\end{array}
\right.$
}
\end{eqnarray}
The previous equation defines the interval
depicting the `area' between 
$\left\lbrace min_{k}, \mu_{ik}-\sigma_{ik} \right\rbrace$
and $\left\lbrace \max_{k}, \mu_{ik}+\sigma_{ik} \right\rbrace$.
For $h$, we adopt 
the Quasi-arithmetic mean \cite{quasi},
i.e.,
\[
r_{i} = \left( \frac{1}{L} \sum_{k=1}^{L} \left(\psi_{ik} \right)^{\alpha} \right)^{1-\alpha}
\]
Based on $\alpha$,
we can alter the results derived by the Quasi-arithmetic 
mean, e.g., if $\alpha = 1$, $r_{i}$ is calculated based on
a `simple' mean.

\section{The Ensemble Learning Scheme}
\label{section5}
We propose the use of a \textit{meta-ensemble learning scheme} that will deliver the final allocation of the 
incoming queries on top of multiple other ensemble schemes. 
Our ensemble scheme tries to `optimize'
the allocation of every $q_{t}$
to the available ENs.
Let $g()$ be a function that results 
the matching degree between 
$q_{t}$ and an EN. The higher the
$g()$ realization is, the higher the matching becomes.
Our selection mechanism deals with the maximization of $g()$
based on queries and ENs characteristics.
In terms of the description of an optimization problem 
that fits in our scenario, we can denote that
our ensemble mechanism tries to 
maximize $g(\mathbf{w}, \{\mathbf{v}\}_{i=1}^{N})$
subject to our models' results 
for delivering 
$l_{i}$, $s_{i}$, $\mathbf{q^{s}}$ and $r_{i}$.
Actually, our ensemble model is responsible to detect 
the most efficient allocations over a complicated
`reasoning' mechanism that adopts multiple ML schemes.

Any ensemble method aims at the creation of 
multiple learning models combined
to deliver the final decision.
The ensemble approach
manages to provide better results
when adopted for prediction
being compared with other
incremental techniques
\cite{ade}.
Some approaches in ensemble learning are
\cite{blanchnik}:
(i) Voting; (ii) Stacking;
(iii) Bagging; (iv) Boosting.
To have the successful application of the ensemble learning model,
we have to apply diversity 
of the obtained results.
The required diversity is achieved by 
\cite{blanchnik}:
(i) the diversity of the models, i.e., 
we have to use different algorithms or 
the same algorithms with different 
parameters;
(ii) the diversity of data, i.e., 
the training data should be different for each of the adopted models.
The most critical issue is the combination of the results. 
Ensemble schemes attract the attention of researchers working in the machine learning domain
based on the view that ensembles are often much more accurate than the individual 
classifiers \cite{dzeroski}. 

\subsection{The Adopted Classifiers}
In our research, we rely on a wide range of classifiers.
and build on the result of already 
defined ensemble schemes to provide our 
meta-ensemble model.
We adopt widely known ensemble 
learning models to provide a more powerful learner.
We adopt the three `basic' ensemble models, i.e.,
(i) the AdaBoost model $Y_{1}$;
(ii) the Stacking model $Y_{2}$;
(iii) the Bagging method $Y_{3}$.
Our meta-ensemble scheme receives the results
of the aforementioned `sub-ensemble' models 
(i.e., $Y_{1}, Y_{2}, Y_{3}$) and delivers the final allocation decision
$Y$.
It should be noted that the aforementioned individual 
ensemble schemes are already focusing on combining
the individual, `single' learners, thus, we 
deliver another, additional, layer of processing.
As the aforementioned schemes are `high-level' ensemble models,
they can be applied over any individual learning algorithm.
Below, we present the list of the adopted individual learners
incorporating into our 
`reasoning' mechanism different types of schemes,
e.g., decision trees, probabilistic models, neural networks.
This exhibits the ability of our mechanism to be unbounded 
of the type of individual learners
and its strength to combine 
various outcomes for the discussed allocations.
Our meta-ensemble model performs a two-level aggregation 
of the outputs retrieved by the individual schemes when applied
over the characteristics of queries and ENs trying to derive the most efficient allocations.
We have to notice that the outputs of the adopted models 
should be in the same interval to be easily aggregated for 
concluding the final decision.

In \cite{re}, the interested reader can study the theoretical background
behind the adoption of ensemble machine learning schemes.
In short, there are three main theories that explain 
the effectiveness of ensemble models.
The first considers ensemble models under the perspective of 
of large margin classifiers \cite{mason}. 
Characteristics of such schemes are 
the enlargement of margins, 
enhancements of the generalization capabilities of Output Coding \cite{allwein} 
and boosting based ensemble algorithms \cite{shapire}. 
The second theoretical approach deals with study of the classical bias/variance decomposition of the error \cite{german}, 
showing that ensembles can reduce variance \cite{breiman}, \cite{lam} or both
bias and variance \cite{kong}, \cite{breiman2}. 
Finally, the third theory is a stochastic discrimination theory focusing on a set-theoretic abstraction to remove all the algorithmic details \cite{kleinberg}.
Adopting such an abstraction, we are able to 
`see' the classifiers as a combination of subsets of points of the feature space and their decisions are also point sets. 
The set of classifiers are, then, just a sample into the power set of the feature space.  

The Adaptive Boosting (AdaBoost) model \cite{freund}
conveys the basic processing of the Boosting model
while combining, in an adaptive way, multiple base learners.
Initially, the Boosting model
builds a first, weak classifier and,
accordingly, a succession of models are built iteratively, 
each one being trained
on a dataset in which points misclassified by the previous model 
are given more weight.
All the adopted models are weighted by their success 
to provide an accurate classification result and 
their outputs are combined through voting.
In any case, the same training dataset is used over and over again. 
The Stacked Generalization (Stacking) model
\cite{wolpert} is another meta-learner
that aims at combining models of different 
types.
Initially, it splits the training dataset in two
disjoint sets and, then, 
trains several base learners on the first part of the dataset.
Accordingly, it tests the base learners with the 
second part of the training dataset.
Afterwards,
it gets predictions in the testing phase 
as the input and correct responses as the output 
to train a higher level learner.
The Bootstrap Aggregation (Bagging) \cite{breiman}
aims at incorporating multiple versions
of the training set through the 
use of sampling with replacement.
Every produced dataset is adopted to train
a different leaning model.
The final output is delivered by averaging the
individual outputs or voting 
when the final results involves a classification process.
Bagging is efficient only when using unstable non-linear models 
because a small change 
in the training set can cause a significant change in the model.

In the aforementioned meta-learning schemes,
we incorporate the following learning algorithms/base learners (to have the necessary diversity in the selection of
algorithms):
(i) C4.5 decision tree;
(ii) Random tree;
(iii) Naive Bayes model;
(iv) Bayesian Network;
(v) Multinomial Naive Bayes model;
(vi) Random Forest;
(vii) Logistic Model Tree;
(viii) REPTree model;
(ix) JRip algorithm;
(x) Multilayer Perceptron.

\subsection{Multi-Classifier Decision Fusion}
Our meta-ensemble model is adopted to realize the aforementioned 
function $g()$, i.e., to deliver the matching and a ranking of the 
available ENs for a query $q_{t}$. 
The proposed model gets as input ENs and $q_{t}$'s characteristics 
exposed by 
$\mathbf{w}$ and $\{\mathbf{v}\}_{i=1}^{N})$
and results the most efficient allocation.
The efficiency of the allocation is realized by the 
`appropriateness' of ENs
to host $q_{t}$ and provide the outcome in the minimum time 
with the best possible performance.
The performance is depicted by the 
matching between $q_{t}$'s constraints (i.e., 
$\mathbf{w}$) and the available datasets 
as well as the ability of ENs to 
minimize the response time (i.e., the execution of $q_{t}$ should start immediately - we target to a low load - while being completed as soon as possible - we target to a high processing speed). 
For detecting the most efficient assignment, we 
adopt the aforementioned ensemble schemes, we get their `opinion'
for every potential allocation and combine them 
through the proposed meta-ensemble model to conclude the final decision.

The meta-ensemble learning scheme is based on 
a training dataset $\mathcal{TD}$ where tuples of context vectors 
are present.
Each training tuple is accompanied by the appropriate 
decision for the `virtual' node that represents.
More formally, every training tuple has the form
$v^{D} = \left\langle o^{TD}, a^{TD}, r^{TD}, l^{TD}, s^{TD}, B \right\rangle$.
The first part of a training tuple is related to the aforementioned context vectors 
while the second part (i.e., $B$)
is related to the final classification result.
We consider a binary classification 
setup with two classes, i.e.,
$B^{1} = 1:$ \textit{ Allocate} and $B^{2} = \overline{B^{1}} = 0 :$ \textit{ Do not allocate}.

The three aforementioned ensemble schemes, i.e.,
the AdaBoost, the Stacking and the Bagging methods are trained 
based on $\mathcal{TD}$ and are adopted to generate the envisioned results
$Y_{1}, Y_{2}, Y_{3}$.
When $Y_{1}, Y_{2}, Y_{3}$ are produced, we have to combine them
to get the final result $Y$ based on which, the final decision is concluded.
In this effort,
we adopt three approaches.
The first is a very `strict' scheme to produce $Y$
relying on the Boolean model originated
in Information Retrieval \cite{manning}.
Based on this model,
the final result is delivered through a simple conjunction, i.e.,
$Y = \prod_{i=1}^{3} Y_{i}$.
This means that if at least one of the aforementioned 
schemes `votes' against the allocation of $q_{t}$ to a specific 
EN/QP, the final result will be the class $B^{2}$. 
To conclude the class $B^{1}$, all the classifiers should agree 
upon this decision.
The second approach is a majority voting scheme where
the majority of classifiers conclude the final result.
Hence, when $count\left\lbrace Y_{i} = B^{1} \right\rbrace$ is at least 2.0,
$Y = B^{1}$ as well.
Our future research plans involve the definition of a more complex
methodology for combining the `opinion'
of the adopted ensemble classification schemes.
In \cite{galar}, the interested reader can 
find a set of aggregation 
techniques for classifiers outputs.
It should be noted that our decision to incorporate the `strict' technique together with the majority voting scheme deals with our intention to use
two representative schemes that are capable of delivering the final result in limited time
requiring limited resources (no additional data structures and variables that consume memory).

Each tuple in $\mathcal{TD}$ represents a combination of
the adopted parameters involving queries and processors characteristics.
For the delivery of the ENs where $q_{t}$ will be allocated, 
we apply the one-over-all (OVA) methodology \cite{han}.
It consists of a widely adopted technique
for multiclass classification with high performance. 
In \cite{rifkin},
a number of optimization algorithms are
compared with the OVA model.
Although these approaches may have a theoretical
interest, it does not appear that they
offer any advantage over a simple OVA scheme.
In addition, OVA compared to other techniques does not 
require any additional time for training.
In our scenario, $N$ classes are available;
one for each EN/QP.
Hence, we adopt $N$ binary classifiers built by the above described 
meta-ensemble model.
Actually, we apply $N$ times the same
meta-ensemble binary classifier for deciding in 
which nodes $q_{t}$ could be allocated.
The classification of an unknown vector 
$v^{D}$ concerns a voting scheme.
If the adopted classifier positively predicts the $i$th node for 
$q_{t}$,
the $i$th EN gets a vote.
If the result is negative (i.e., class $B^{2}$) all the remaining 
nodes except $i$ get a vote. 
The node(s) with the most votes is selected
to host $q_{t}$. In the case of ties,
we rank the nodes based on their load.

\section{Experimental Evaluation}
\label{section6}
We perform a high number of simulations adopting synthetic traces and 
datasets found in the literature.
In addition, 
we compare the proposed scheme with our previous work.
The aim is to reveal the advantages and the drawbacks of our model
when adopted for the allocation of analytics queries.
In the upcoming sub-sections, we describe our
experimental setup and the assessment of our model.

\subsection{Performance Metrics \& Experimental Setup}
We perform our experiments with a custom software 
written in Java that contains a number of classes for the 
simulation of the adopted nodes and datasets present in them.
Initially, 
we deal with the accuracy of the proposed
FCP for detecting the correct complexity class of 
a query.
For this, we define the metric $\upsilon$ as follows:
\begin{eqnarray}
\upsilon = \frac{Q_{C}}{|Q|}
\label{rp:perf0}
\end{eqnarray}
where $Q_{C}$ is the number of the correct predictions.
Obviously, we want $\upsilon \to 1$ to enjoy the best possible performance.
We also focus on the time required for the allocation of each incoming 
query that affects the final throughput of $QC$s.
We define the metric $R$ that 
represents the discussed throughput, i.e., 
\begin{eqnarray}
R = \frac{|Q|}{\sum_{i=1}^{|Q|} T^{c}_{i}}
\label{rp:perf1}
\end{eqnarray}
where $T^{c}_{i}$ is the conclusion time (in ms) 
for the $i$th query. 
The \textit{conclusion time} is the time required to
allocate a query to one of the available nodes just after its reception. 
$R$ depicts the amount of queries executed by 
a QC in a time unit (i.e., ms).
The higher the $R$ is, the higher the performance
of the proposed model becomes.

We also adopt parameters $l^{*}$ and $s^{*}$ depicting the load and the speed of the
selected node, respectively.
Based on these parameters, we define 
the metrics $D_{l}$ and $D_{s}$.
Both of them represent the `distance' of $l$ and $s$ of the selected node 
from the optimal load and speed in the entire group.
As the optimal load, we define the lowest load ($l_{min}$) observed in the available nodes at the decision time.
In a similar way, as the optimal speed, we define the highest speed ($s_{max}$) among the available nodes
at the decision time.
Through the adopted performance metrics, we try to reveal if the proposed model
is capable of selecting the best possible node at the decision time.
If this is true, our scheme allocates the incoming queries to nodes that 
will return the final response in the minimum time.
The following equations hold true:
\begin{eqnarray}
D_{l} = l^{*} - l_{min}
\label{rp:perf2}
\end{eqnarray}

\begin{eqnarray}
D_{s} = s_{max} - s^{*}
\label{rp:perf3}
\end{eqnarray}

We compare the proposed Conjunctive Scheme $CS$ and the
Majority Voting Scheme ($MVS$) with our previous work 
\cite{kolomvatsos2}, \cite{kolomvatsos3}, \cite{kolomvatsos1}
(Model 1 - M1, Model 2 - M2, Model 3 - M3, respectively).
We have already discussed the novelty of the current framework in contrast to our previous schemes,
thus, for providing a holistic comparison in our experimental evaluation, we deliver numerical results.
We evaluate our model for different 
realizations of $N$ 
getting $N \in \left\lbrace 10, 50, 100, 500 \right\rbrace$.
For each experiment, we consider the 
execution of 1,000 queries and 
take results for the aforementioned models.
For the type of the incoming queries and
the delivery of their complexity class, we rely on the 
dataset and the methodology presented in \cite{kolomvatsos4}.
For the remaining parameters, i.e., query vectors, $a$, $l$, $s$ and the 
data reported in nodes, we rely on a set of traces.
Our data are retrieved by the following traces:
\begin{itemize}
	\item Dataset 1. A synthetic trace based on the Uniform distribution defining $a$ for each query. For the generation of each value, we consider an maximum $a$ equal to 10.
	\item Dataset 2. A synthetic trace based on the Uniform distribution delivering values for query vectors, the data present in each node, $l$ and $s$. 
	\item Dataset 3. A synthetic trace based on the Gaussian distribution delivering values for query vectors, the data present in each node, $l$ and $s$.
	\item Dataset 4. The trace reported by \cite{akay}. From this dataset, we adopt the processor utilization dimension that depicts the percent of time that threads are running in a processor. The data are incorporated to our simulator for providing the realizations of $l$. The remaining parameters take values as described for Datasets 2 \& 3.
\end{itemize}

\subsection{Performance Evaluation}
Initially, 
we report on the accuracy of the proposed
FCP for delivering $\mathbf{q^{s}}$.
Recall that values present in $\mathbf{q^{s}}$ depict the 
`membership' of each query to the available complexity classes.
Executing the FCP for the queries dataset reported in \cite{kolomvatsos4},
we get $\upsilon = 1.0$ for a threshold less than or equal to 0.8.
Over this threshold, we consider that the query has the complexity 
of the class corresponding to the index of the discussed value in $\mathbf{q^{s}}$.
As already described, it is difficult to assign an individual complexity class for each query,
thus, the maximum value of $\mathbf{q^{s}}$ represents the class that 
exhibits the highest similarity with $q_{t}$.
If we get the threshold equal to 0.9, $\upsilon \to 0.55$.
It is worth noticing that the maximum value in $\mathbf{q^{s}}$ in the entire set of 
our experiments is over 0.8.

\begin{figure}[H]
  \centering
    \includegraphics[width=0.85\textwidth]{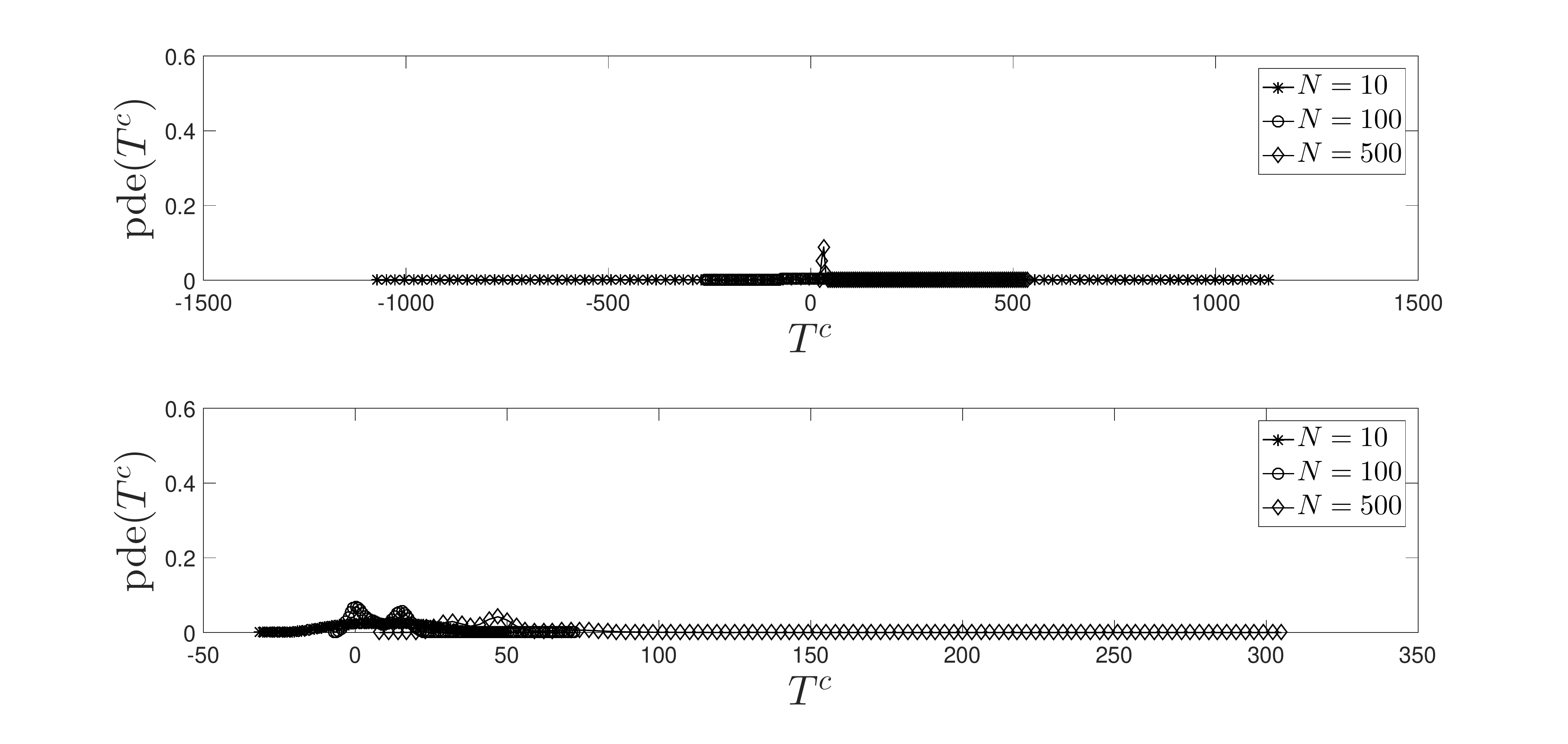}  
   \caption{Pde of the allocation time for CS (up: Uniform distribution, down: Gaussian distribution).}
\label{figure01}
\end{figure}

\begin{figure}[H]
  \centering
    \includegraphics[width=0.85\textwidth]{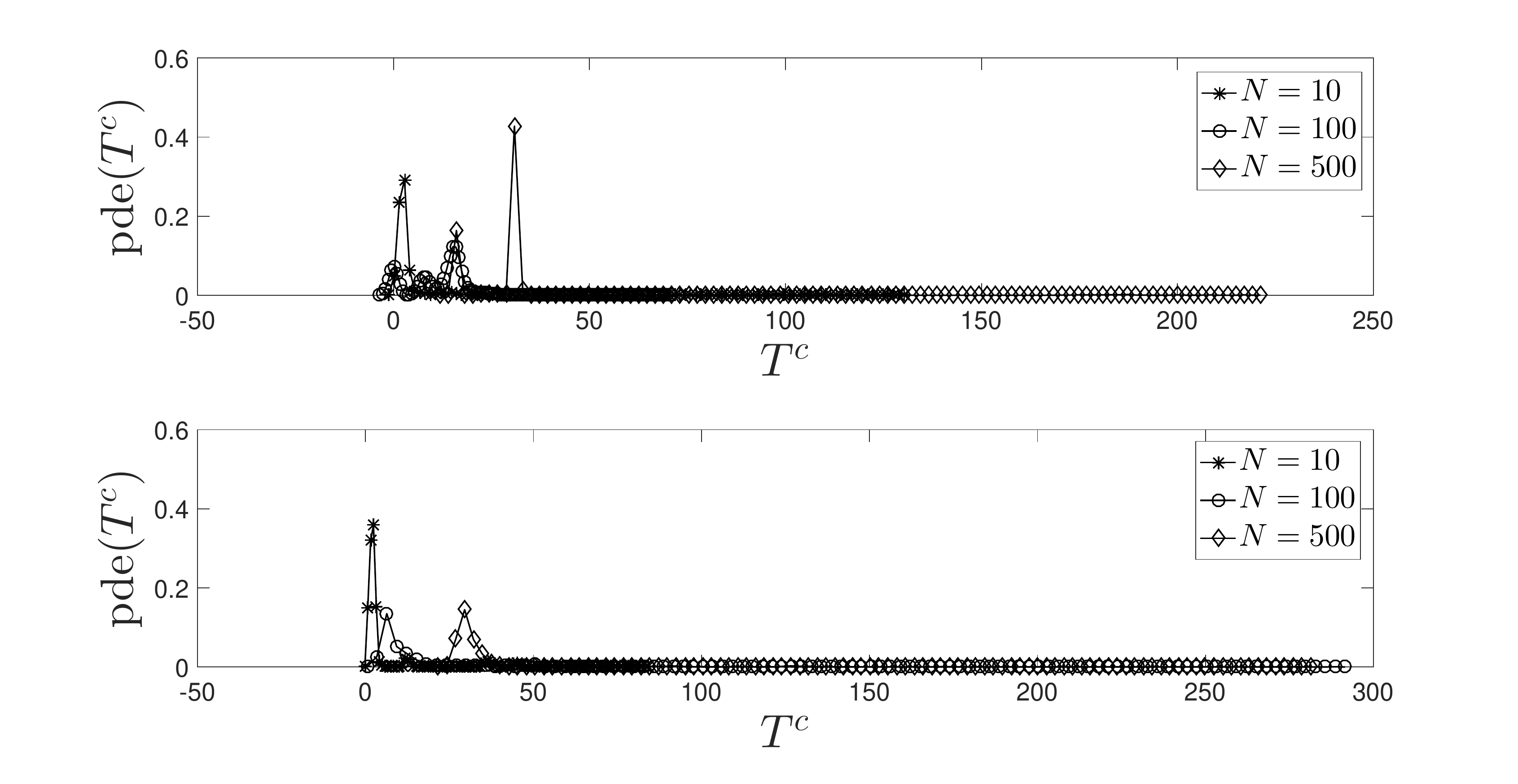}
\caption{Pde of the allocation time for MVS (up: Uniform distribution, down: Gaussian distribution).}
\label{figure02}
\end{figure}

We report on the probability density estimate ($pde$) of the adopted performance metrics achieved 
by the proposed model.
In Figure \ref{figure01},
we see our results for $T^{c}$ (the time required in ms
for the allocation of a query).
We observe that the uniformity of the 
data (upper figure) leads to the uniformity of the
discussed performance metric.
In contrast, the use of the Gaussian distribution
(lower figure) leads to results affected by the 
number of the available ENs.
In the latter case, 
a low number of ENs (e.g., 
$N << 500$) leads to a low $T^{c}$.
In these scenarios, the throughput of QCs is high
managing to allocate multiple queries in a limited time interval.
It should be noted that the discussed results refer in 
the $CS$.
If we pay attention on the $MVS$ (see Figure \ref{figure02}), we observe the uniformity 
of $T^{c}$ as well.
However, the higher the $N$ is the higher the $T^{c}$
becomes.
This is natural, as QCs should process multiple context 
vectors as inputs into the matching process.
In any case, if we consider that the provided results are in the scale of ms, we can conclude that the proposed mechanism, even if it involves multiple classification models, manages to deliver the final allocation in 
a limited time interval.

\begin{figure}[H]
  \centering
    \includegraphics[width=0.85\textwidth]{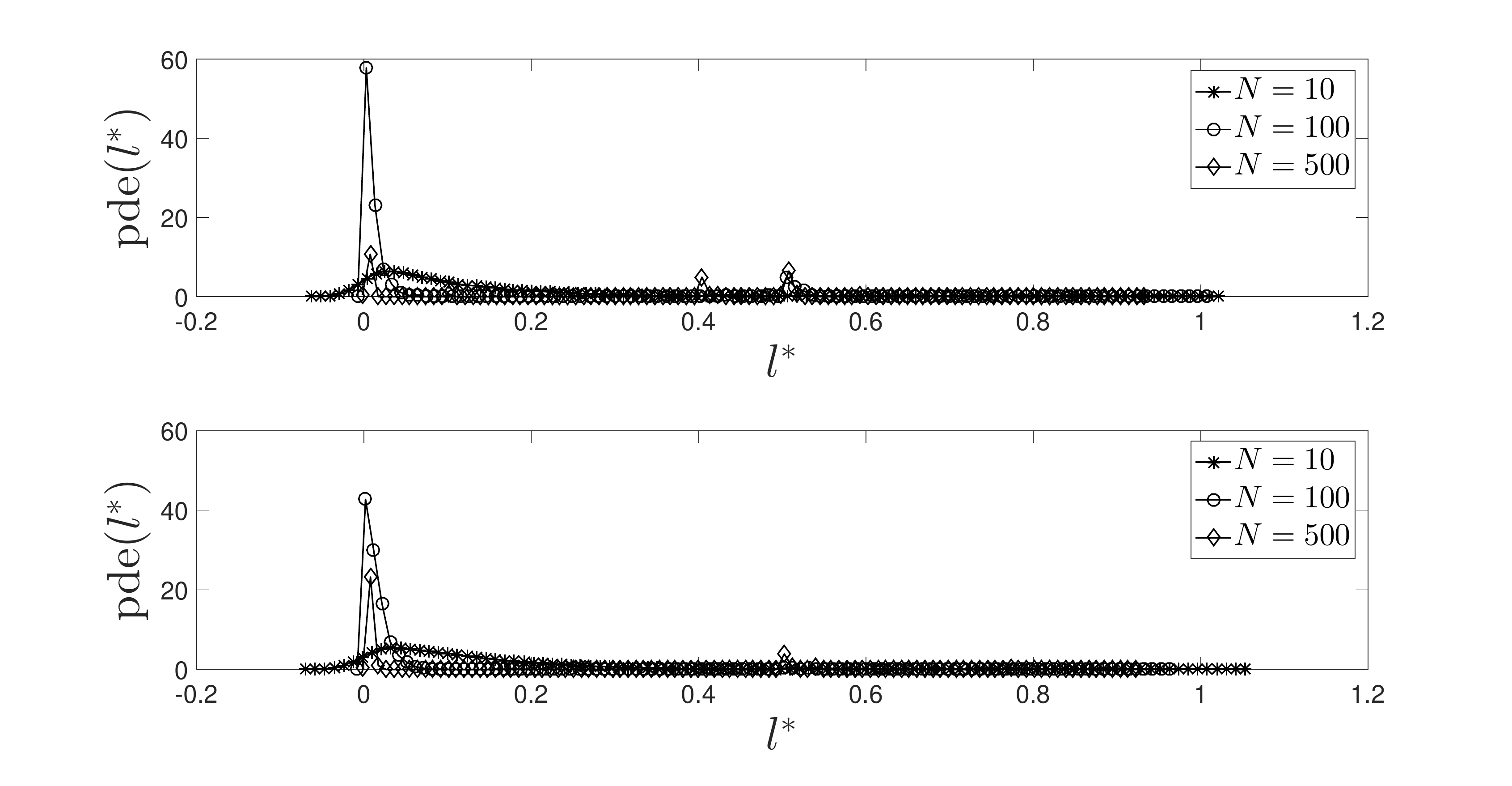}
\caption{Pde of $l^{*}$ for CS (up: Uniform distribution, down: Gaussian distribution).}
\label{figure03}
\end{figure}

\begin{figure}[H]
  \centering
    \includegraphics[width=0.85\textwidth]{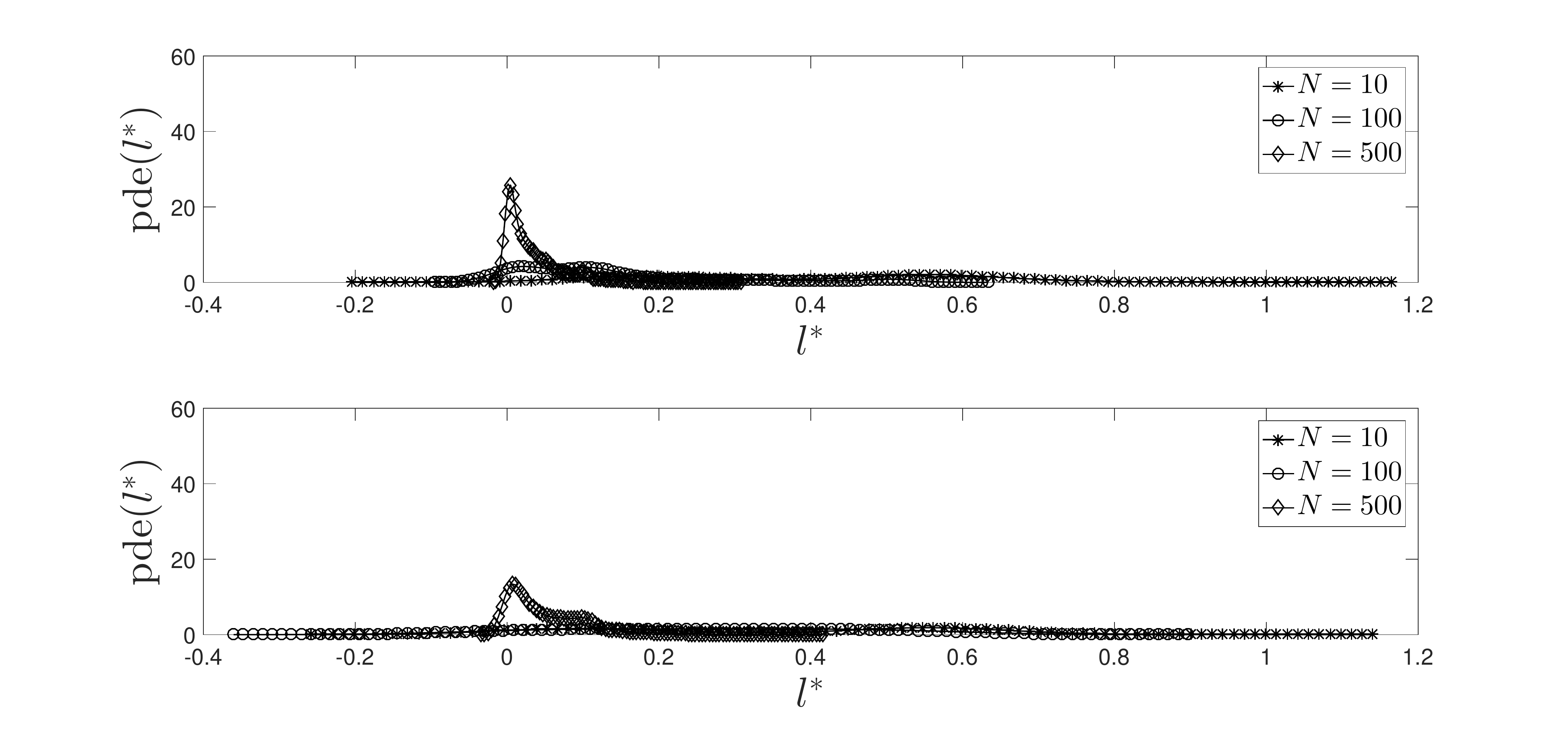}
\caption{Pde of $l^{*}$ for MVS (up: Uniform distribution, down: Gaussian distribution).}
\label{figure04}
\end{figure}

In Figures \ref{figure03} \& \ref{figure04},
we present our results related to $l^{*}$ for both
$CS$ and $MVS$.
The higher the $N$ is, the lower the $l^{*}$ becomes.
This is more intense in $MVS$.
A high number of ENs give us more opportunities to 
seek for the best possible node.
The provided classification scheme and the OVA approach
manage to conclude nodes that exhibit low load, thus,
it could deliver the final result in a limited time
(however, this finally depends on the complexity of the process demanded by the query).
A low number of ENs may lead to a higher load compared to the previous scenario in both models, i.e., $CS$, $MVS$.

\begin{figure}[H]
  \centering
    \includegraphics[width=0.85\textwidth]{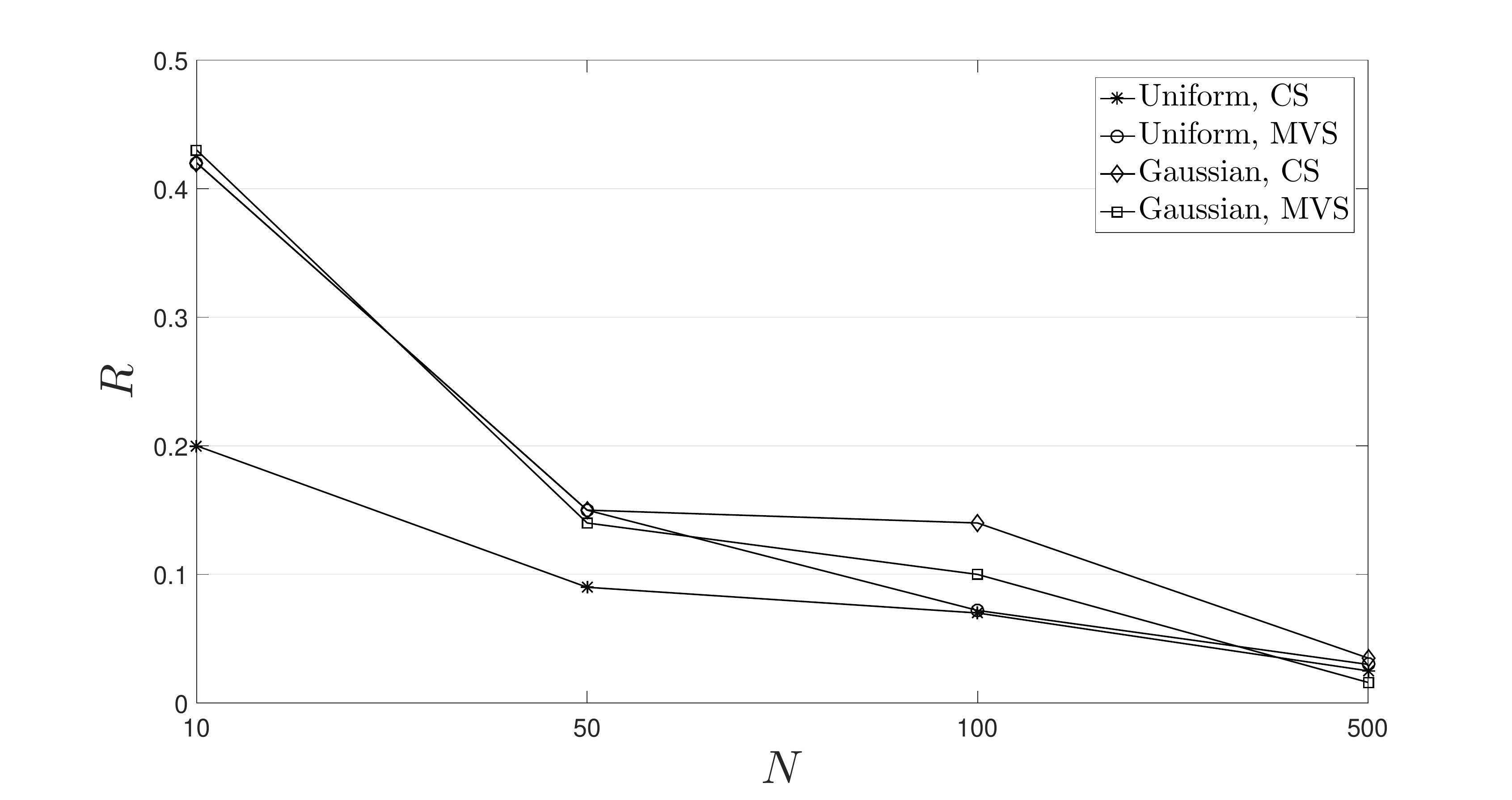}
\caption{The throughput of the proposed model.}
\label{figure05}
\end{figure}

In Figure \ref{figure05}, we plot the throughput 
of the proposed mechanisms. The throughput is minimized 
when we have to deal with multiple nodes; a natural consequence of the involvement of multiple processes until the final decision.
The $CS$ exhibits the best performance among the models,
however, the difference between them is limited when
$N \to 500$.
When $N=10$, the $CS$ exhibits the worst performance when
the Uniform distribution is adopted to produce the values for our parameters.
In terms of the number of queries allocated in a second,
we get that the $CS$ can allocate 420 to 35 queries while the $MVS$ can allocate 430 to 16 queries. 

\begin{figure}[H]
  \centering
    \includegraphics[width=0.85\textwidth]{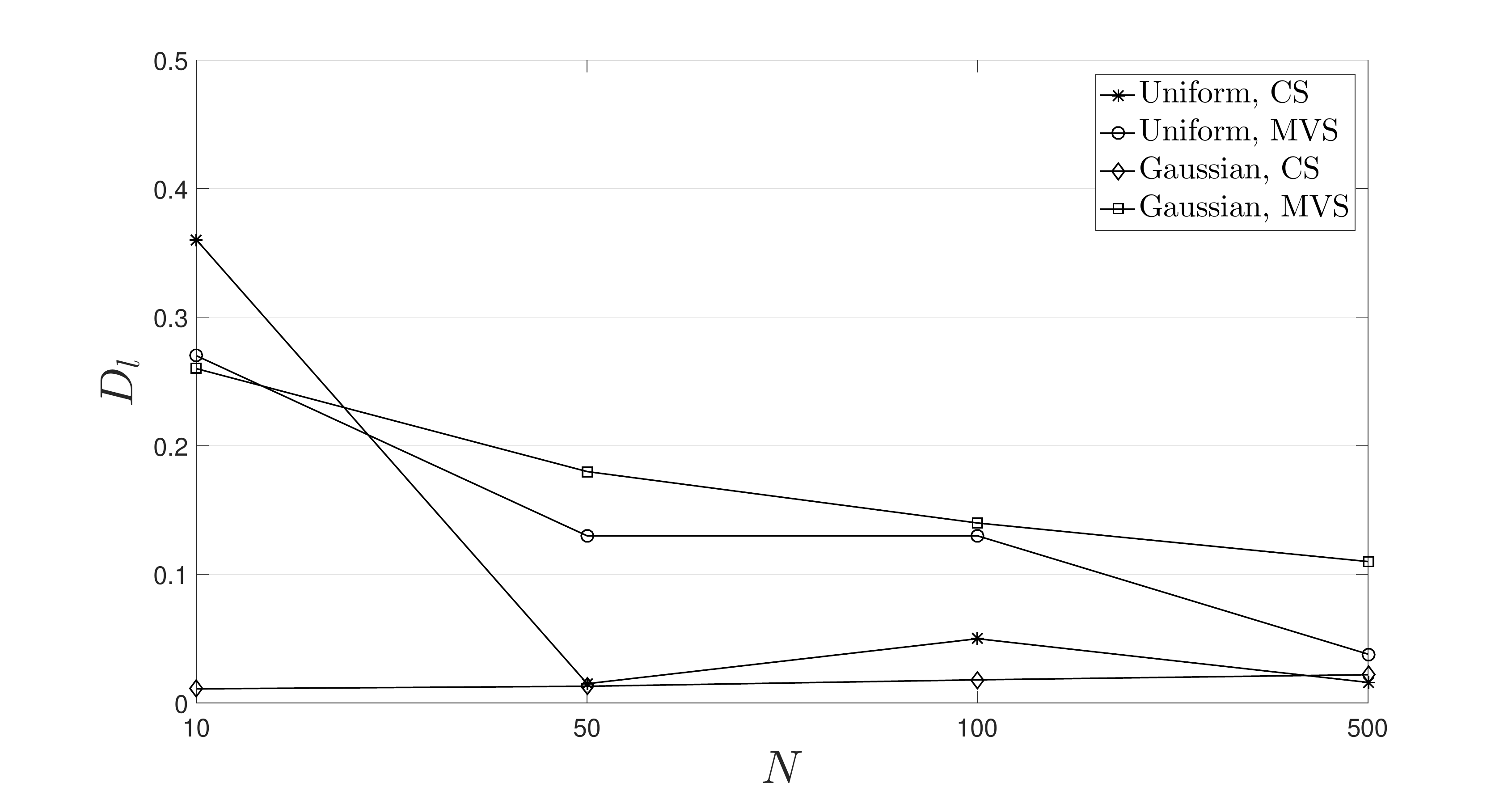}
\caption{The load of the selected node as delivered by the proposed model compared to the lowest available load.}
\label{figure06}
\end{figure}

In Figure \ref{figure06}, we plot our results related to the 
optimality of the proposed model as far as $D_{l}$ concerns.
The $CS$ outperforms the $MVS$ except one experimental 
scenario ($N=10$).
Recall that if at least one of the adopted ensemble schemes rejects the allocation, the query is not classified to the corresponding ENs.
This means that the $CS$ tries to take a decision under unanimity that is not the case in the $MVS$.
The significant is that as $N \to 500$, both models
result in a limited distance with the optimal solution.
For instance, the $CS$ results in a distance equal to 
0.016 \& 0.022 for the Uniform and the Gaussian distributions, respectively.
The $MVS$ results in a distance equal to 
0.038 \& 0.110 for the Uniform and the Gaussian distributions, respectively.
In any case, these numbers 
exhibit the efficiency of the proposed scheme 
keeping in mind that we plot the average 
$D_{l}$ throughout our experiments.

\begin{figure}[H]
  \centering
    \includegraphics[width=0.85\textwidth]{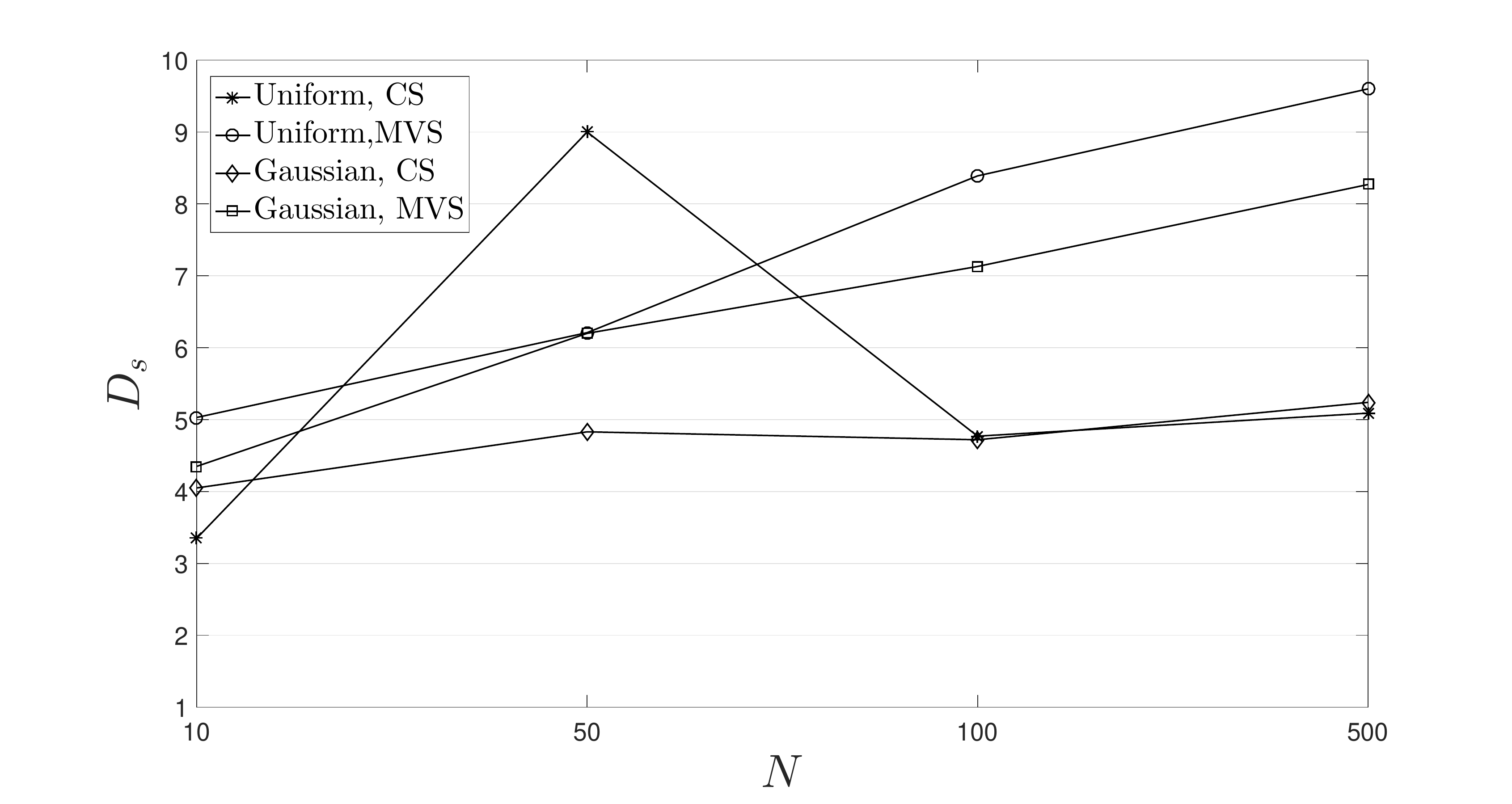}
\caption{The speed of the selected node as delivered by the proposed model compared with the highest available speed.}
\label{figure07}
\end{figure}

The next set of our experiments deal with the 
speed of the selected node.
We should note that the speed of processors is randomly selected as depicted by the aforementioned traces with a maximum value equal to 10.
In Figure \ref{figure07}, we present the relevant results.
The distance from the optimal speed in the group increases as $N$ increases.
Especially when $N=500$, the distance is high.
We have to compare these results with the 
results derived for $D_{l}$.
The increased number of nodes positively affects the
load of the selected node while it negatively affects the speed of the selected EN.
The proposed model pays more attention
on the minimization of load, however, 
we should note that this is affected by the training 
set adopted to prepare the ensemble classifiers for 
decision making.
This dependence on the training set is `typical'
for any supervised machine learning algorithm, thus, 
in our future research plans we want to build an ensemble
model based on unsupervised techniques and compare the performance.

\begin{figure}[H]
  \centering
    \includegraphics[width=0.85\textwidth]{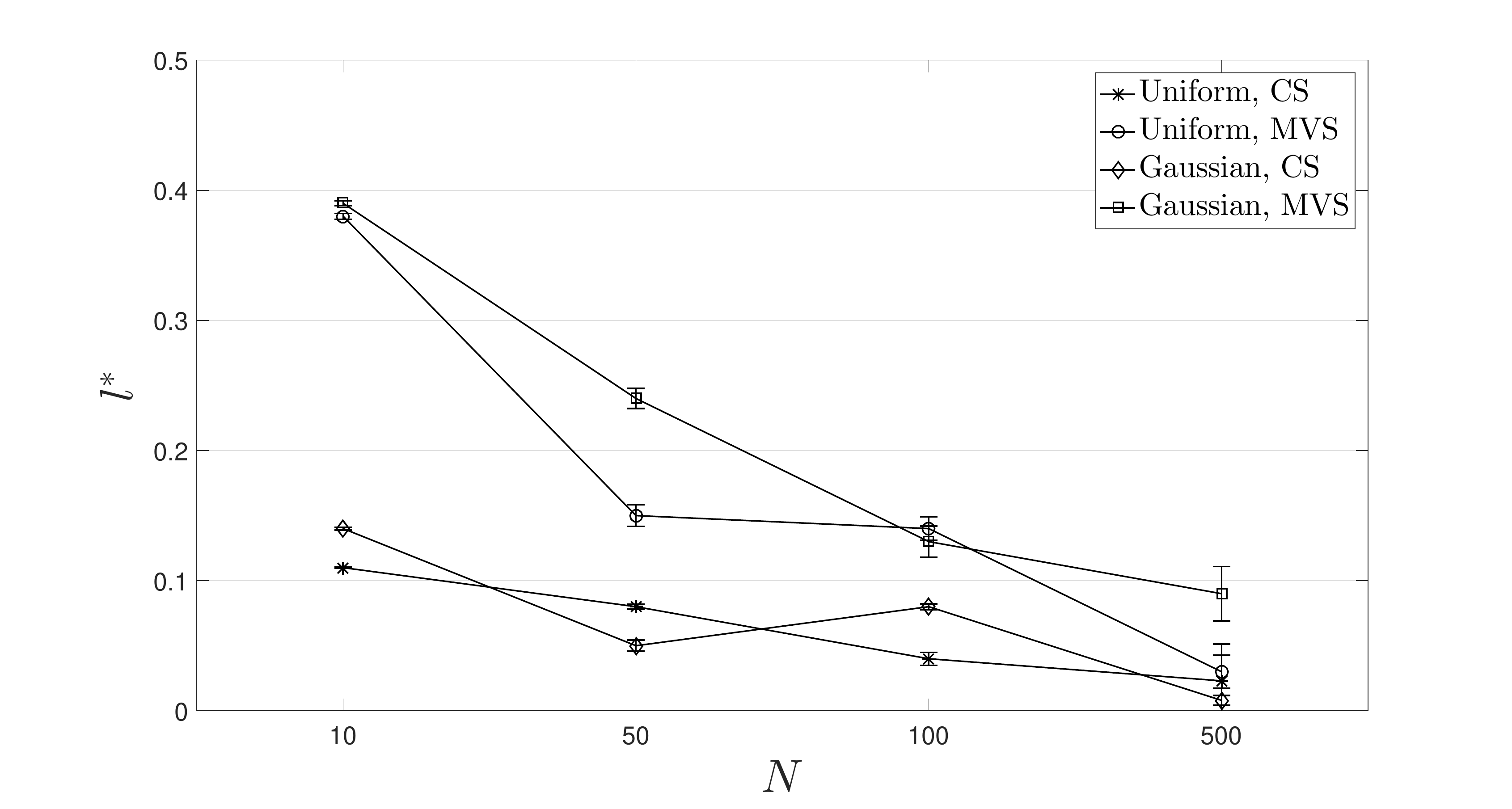}
  \caption{The load of the selected node as delivered by the proposed model.}
\label{figure08}
\end{figure}

In Figure \ref{figure08}, we plot the $l^{*}$ realizations, i.e., the average load of the selected nodes
for each of the 1,000 queries.
The load is limited for a high number of nodes, which means that the burden added to ENs is minimized. 
The interesting is that the statistical error in our measurements is also limited (for the majority of the scenarios) exhibiting the stability of our model.
In this set of our experiments, the $CS$ exhibits the best performance compared to the $MVS$.

\begin{figure}[H]
  \centering
    \includegraphics[width=0.85\textwidth]{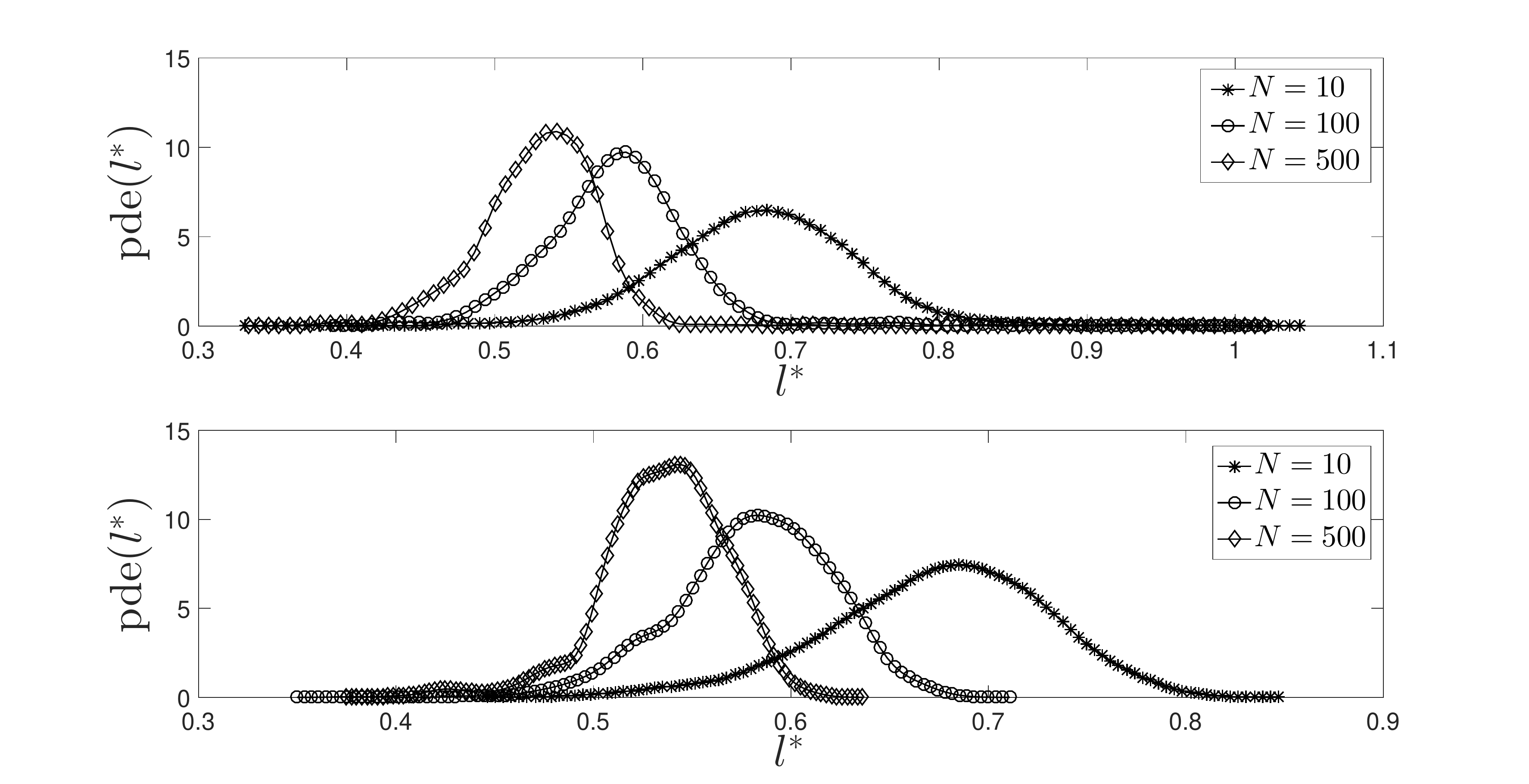}
  \caption{Pde for the load of the selected node based on Dataset 4 (up: CS, down: MVS).}
\label{figure09}
\end{figure}

\begin{figure}[H]
  \centering
    \includegraphics[width=0.85\textwidth]{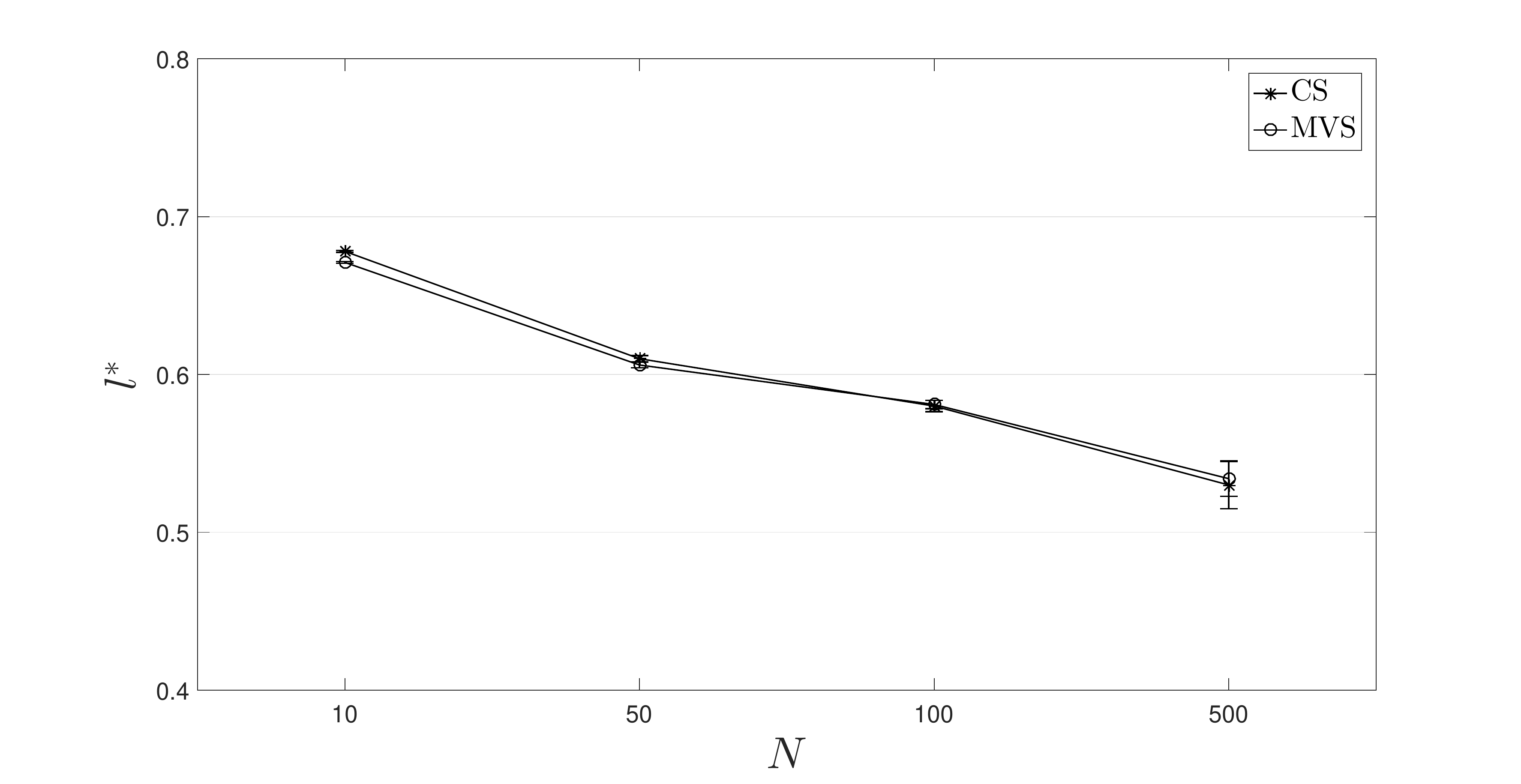}
  \caption{The load of the selected node as delivered by the proposed model based on the real trace.}
\label{figure10}
\end{figure}

The next set of experiments deals with the dataset 
provided by \cite{akay}.
Our results are presented in Figures \ref{figure09}
\& \ref{figure10}.
The outcomes confirm our previous observations 
when adopted the synthetic traces.
An increment in the number of ENs positively affects the
performance of the model and leads to the selection of nodes
with a low load.
The average load of the finally selected node is similar 
for $CS$ and $MVS$.

\begin{figure}[H]
  \centering
    \includegraphics[width=0.85\textwidth]{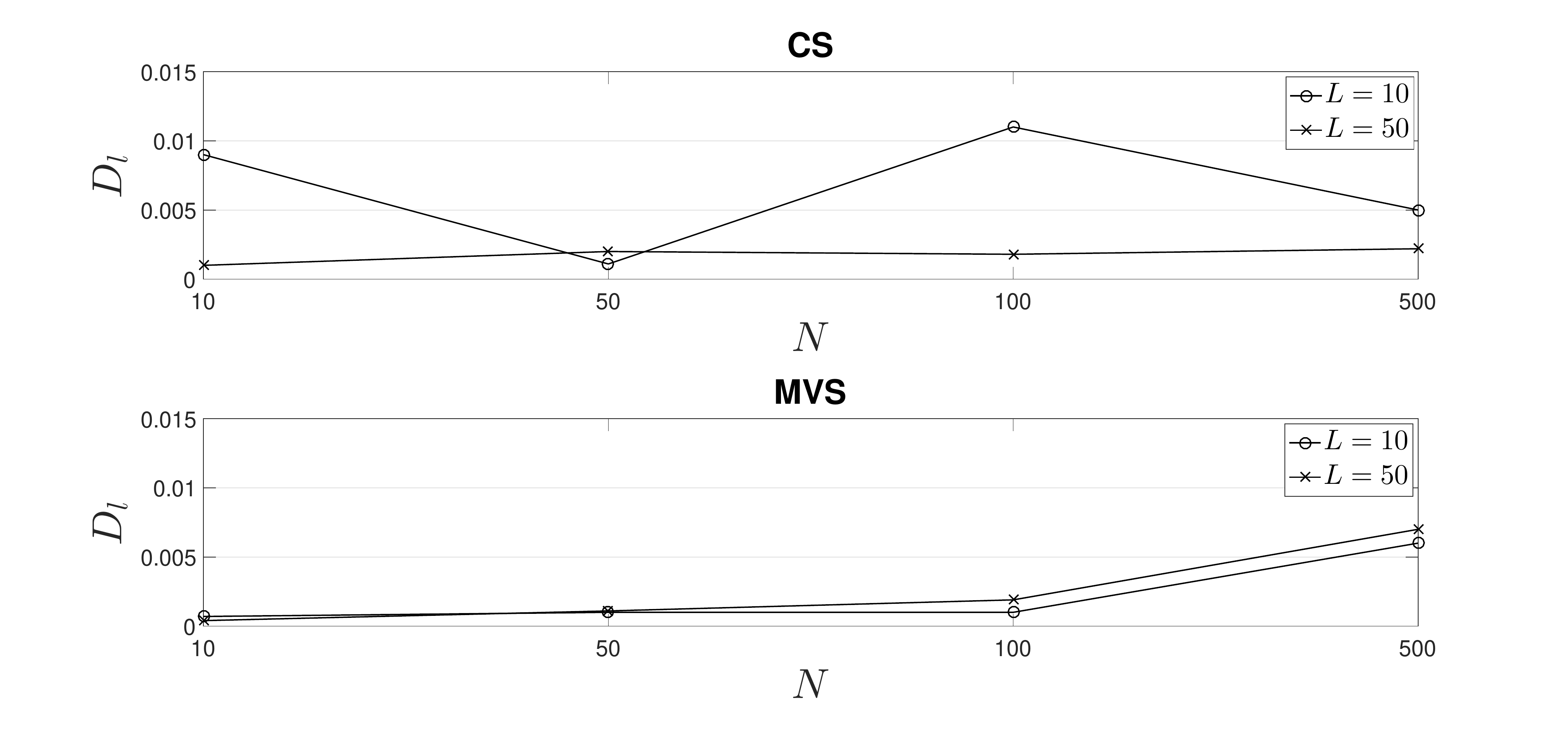}
  \caption{Our sensitivity results for various $L$ realizations.}
\label{figure12}
\end{figure}

\begin{figure}[H]
  \centering
    \includegraphics[width=0.85\textwidth]{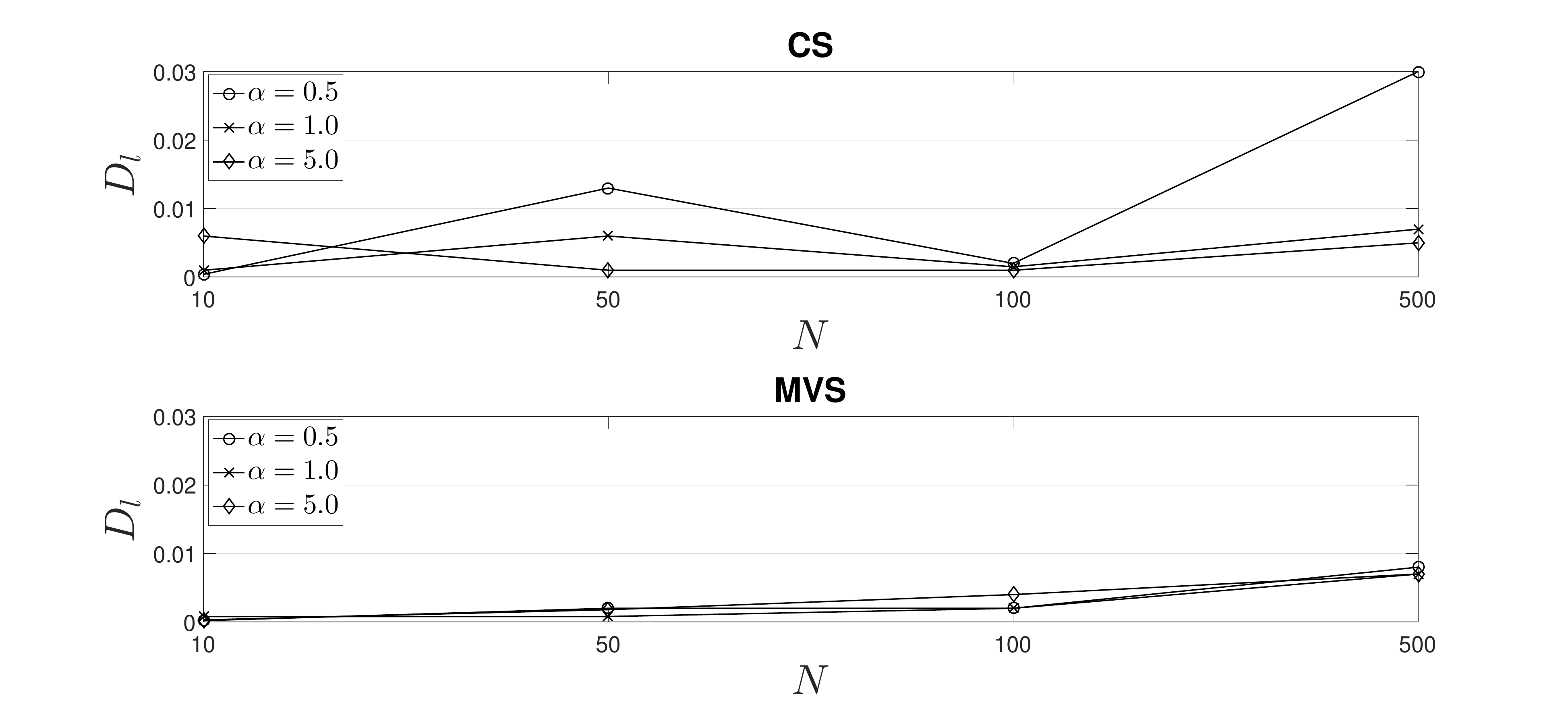}
  \caption{Our sensitivity results for various $\alpha$ realizations.}
\label{figure11}
\end{figure}

We also report on our model's sensitivity in parameters $L$ and $\alpha$.
Recall that $L$ depicts the number of dimensions into 
the available datasets and $\alpha$ is adopted to `smooth' the result of the Quasi arithmetic mean 
when we calculate $\mathbf{q}^{s}$ and $r_{i}$.
In Figure \ref{figure11}, we consider Dataset 4 and  
$L \in \left\lbrace 10, 50\right\rbrace$ to present our results for $D_{l}$.
We observe that the difference with the optimal 
(minimum) load increases as $N$ increases, however,
$D_{l}$ is below 0.015 for both, i.e., 
the $CS$ and the $MVS$.
Apart from that, we can conclude that $L$ does not 
heavily affect the performance of the proposed approach.
Our mechanism can be adopted for a high number of dimensions
to result the best possible node to host the incoming queries.
It is worth noticing that an increased $L$ will lead to an increased processing time affecting $R$.
In Figure \ref{figure12}, we present our results
for $\alpha \in \left\lbrace 0.5, 1.0, 5.0 \right\rbrace$ and for Dataset 4.
Again, the proposed scheme is not heavily affected by the $\alpha$ realization exhibiting efficiency in the detection of the best possible node to host the incoming queries.
The majority of results are below 0.01 (the selected node is very close to the node exhibiting the lowest load).

We also compare the proposed model with our previous efforts in the domain.
The comparison refers in the throughput
of QCs as well as the load of the selected node.
Our model results an $R$ 
varying from 
0.42 to 0.022 for 
$N \in \left\lbrace 10, 50, 100, 500\right\rbrace$
and average $l^{*} \in \left\lbrace [0.008, 0.14], [0.03, 0.39] \right\rbrace$, for the $CS$ and the $MVS$, respectively.
The model M1 discussed in 
\cite{kolomvatsos2} adopts an optimal stopping scheme that 
sequentially examines a set of processors before it 
delivers the final allocation.
M1 results $R$ values between 
0.02 (for $N = 50$)
and 0.27 (for $N=10$).
Hence, for a low number of nodes 
$CS$ and $MVS$ outperform M1 while in the remaining
scenarios we observe a similar performance.
The Model M2 presented in 
\cite{kolomvatsos3} adopts a learning mechanism
accompanied by a load balancing module.
The discussed scheme manages to 
allocate 58.48 queries for $N=100$
to 454.55 queries for $N=20$.
The average load of the selected node (the learning scheme)
is around 0.10 for
$N \in \left\lbrace 2,5,20,50,100\right\rbrace$.
The average load for the clustering scheme
is between 0.160 and 0.670.
Our current models outperform the clustering scheme adn exhibit 
a similar performance with the learning model.
Finally, the model M3 discussed in 
\cite{kolomvatsos1} adopts a `simple' learning scheme for the decision making.
The performance of M3 related to the throughput
is 50-100 queries
per time unit when $N \to 500$.
As we observe, our current approach outperforms, in the majority of the scenarios, 
our previous efforts leading to an efficient model.
We have to take into consideration that the current scheme is more complex than the previous ones
as it involves multiple classification decisions before the final result is in place.
In addition, the current model takes every decision
based on multiple parameters providing a holistic solution to the discussed problem.
The throughput is at acceptable levels which means that 
the proposed scheme does not delay to provide the final allocation.
In addition, as depicted by the comparison for the $l$ parameter, the current model manages to result the nodes with a low load and, in many cases, better than our previous efforts.

\section{Conclusions}
\label{section7}
The present status of Web, IoT and pervasive computing
deals with the presence of numerous devices able to collect and process data.
All these data are transferred to the Cloud to be the subject of 
further processing and knowledge production.
Usually, such knowledge has the form of analytics 
that are the response in queries defined by users or applications.
In this paper, we focus on an edge computing scenario where 
the collected data could be processed in edge nodes to reduce the 
latency in the provision of responses. 
Knowledge can be produced locally, close to the source of data
and end users. 
A number of edge nodes can serve as aggregation points of data 
reported by the surrounding devices. 
As data are geo-distributed, a critical questions is raised: 
in which edge node a query should be executed?
Our paper tries to respond to this question and presents 
a model that efficiently allocates the incoming queries to the appropriate 
nodes. 
We provide a decision making mechanism, i.e., a meta-ensemble learning
model responsible to select the best possible node to allocate a query.
The decision is made on top of multiple parameters related to queries as well as 
to the available nodes.
Our model takes into consideration the form of queries,
their complexity, the deadline for having the final response
together with the load and the speed of nodes.
More importantly, we propose the use of a parameter that represents 
the distance between queries and datasets present in nodes.
The aim is to exclude nodes that do not own data related to the 
conditions raised by queries as these nodes cannot 
provide responses that `match' the requirements of the incoming queries.
Our meta-ensemble learning scheme adopts multiple 
ensemble learning models to support a powerful 
decision making mechanism.
We describe the performance of our framework through a large set of 
simulations adopting synthetic traces and traces found in the literature.
Our evaluation results show that our model is capable of 
efficiently allocating the incoming queries towards the support 
of (near) real time responses.   
In the first places of our future research plans is the definition of  
an adaptive learning model that will be fully aligned
with the environment of the edge nodes.  
This is important as it will save 
resources devoted to the training process
through an `incremental' training approach.

\section*{Acknowledgment}
This research received funding from the European's Union Horizon 2020 research and innovation programme under the grant agreement No. 745829.



%

\end{document}